\documentclass[pra,twocolumn,aps,letterpaper, preprintnumbers,superscriptaddress]{revtex4-2}

\usepackage{graphicx}
\usepackage{amsmath,amssymb,amsfonts}
\usepackage{subfigure}
\usepackage{enumerate}
\usepackage{enumitem} 
\usepackage[usenames,dvipsnames]{color}
\usepackage[colorlinks=true,citecolor=blue,urlcolor=black]{hyperref}
\usepackage{tcolorbox}
\usepackage[linesnumbered,algoruled,boxed,lined]{algorithm2e} 
\usepackage{siunitx}
\usepackage{ragged2e}
\usepackage{multirow}
\usepackage{amssymb}

\newtheorem{lemma}{Lemma}

\newtheorem{proof}{Proof}
\newtheorem{definition}{Definition}

\newcommand{\arrange}[2]{A_{#1}^{#2}}

\begin{document}
	
\title{Beating the Fault-Tolerance Bound and Security Loopholes for Byzantine Agreement with a Quantum Solution}
\author{Chen-Xun Weng}\email{These authors contributed equally to this work.}\affiliation{National Laboratory of Solid State Microstructures and School of Physics, Collaborative Innovation Center of Advanced Microstructures, Nanjing University, Nanjing 210093, China}
\author{Rui-Qi Gao}\email{These authors contributed equally to this work.}\affiliation{National Laboratory of Solid State Microstructures and School of Physics, Collaborative Innovation Center of Advanced Microstructures, Nanjing University, Nanjing 210093, China}
\author{Yu Bao}\affiliation{National Laboratory of Solid State Microstructures and School of Physics, Collaborative Innovation Center of Advanced Microstructures, Nanjing University, Nanjing 210093, China}
\author{Bing-Hong Li}\affiliation{National Laboratory of Solid State Microstructures and School of Physics, Collaborative Innovation Center of Advanced Microstructures, Nanjing University, Nanjing 210093, China}
\author{Wen-Bo Liu}\affiliation{National Laboratory of Solid State Microstructures and School of Physics, Collaborative Innovation Center of Advanced Microstructures, Nanjing University, Nanjing 210093, China}
\author{Yuan-Mei Xie}\affiliation{National Laboratory of Solid State Microstructures and School of Physics, Collaborative Innovation Center of Advanced Microstructures, Nanjing University, Nanjing 210093, China}
\author{Yu-Shuo Lu}\affiliation{National Laboratory of Solid State Microstructures and School of Physics, Collaborative Innovation Center of Advanced Microstructures, Nanjing University, Nanjing 210093, China}
\author{Hua-Lei Yin}\email{hlyin@ruc.edu.cn}\affiliation{National Laboratory of Solid State Microstructures and School of Physics, Collaborative Innovation Center of Advanced Microstructures, Nanjing University, Nanjing 210093, China}\affiliation{Department of Physics and Beijing Key Laboratory of Opto-electronic Functional Materials and Micro-nano Devices, Key Laboratory of Quantum State Construction and Manipulation (Ministry of Education), Renmin University of China, Beijing 100872, China}
\author{Zeng-Bing Chen}\email{zbchen@nju.edu.cn}
\affiliation{National Laboratory of Solid State Microstructures and School of Physics, Collaborative Innovation Center of Advanced Microstructures, Nanjing University, Nanjing 210093, China}

\date{\today}
	
\begin{abstract}
Byzantine agreement, the underlying core of blockchain, aims to make every node in a decentralized network reach consensus. Classical Byzantine agreements unavoidably face two major problems. One is $1/3$ fault-tolerance bound, which means that the system to tolerate $f$ malicious players requires at least $3f+1$ players. The other is the security loopholes from its classical cryptography methods. Here, we propose a  Byzantine agreement framework with unconditional security to break this bound with nearly $1/2$ fault tolerance due to multiparty correlation provided by quantum digital signatures. \textcolor{black}{It is intriguing that quantum entanglement is not necessary to break the $1/3$ fault-tolerance bound, and we show that weaker correlation, such as asymmetric relationship of quantum digital signature, can also work.} Our work strictly obeys two Byzantine conditions and can be extended to any number of players without requirements for multiparticle entanglement. We experimentally demonstrate three-party and five-party consensus for a digital ledger. Our work indicates the quantum advantage in terms of consensus problems and suggests an important avenue for quantum blockchain and quantum  consensus networks. 
\end{abstract}

\maketitle 

\section{Introduction}
Byzantine agreement requires solving the fundamental consensus problem initially posed in 1982 known as the Byzantine Generals Problem, which can ensure the smooth functioning of a decentralized system under the attacks of malicious players~\cite{lamport1982byzantine,Extance2015future}. This problem can be translated into a `commanding general-lieutenants' model, where the commanding general is randomly selected from among all the Byzantine generals and the others become lieutenants to reach consensus on the commanding general's order (see Appendix~\ref{appA} for details). For a strict Byzantine agreement, there are two necessary \emph{interactive consistency} (IC) Byzantine conditions. The first is that all loyal lieutenants obey the same order (IC$_1$), and the second is that every loyal lieutenant obeys the order of the commanding general if the commanding general is loyal (IC$_2$). Only when both conditions are satisfied can the system reach consensus. For an $N$-party system, however, classical Byzantine agreement (CBA) protocols~\cite{castro1999practical,Castro2002practical,aublin2013RBFT,miller2016HBBFT,yin2019hotstuff,guo2020dumbo,lu2020dumbo} that tolerate $f$ malicious players require $N\ge 3f+1$ players; namely, the fault-tolerance bound is $1/3$~\cite{pease1980reaching,dolev1986possibility,fischer1986easy,fitzi2001advances}. Thus, the three-party consensus problem is naturally unsolvable for CBA even using the authentication classical channel~\cite{kiktenko2018quantum}. The other issue is the security loopholes of CBA's widely used public-key encryption and one-way hash function~\cite{menezes2018handbook}, which are seriously threatened by quantum computing~\cite{shor1994algorithms,grover1997quantum,arute2019quantum,fedorov2018quantum,wei2020full,Fernandez2020towards,zhou2022experimental,huang2022quantum,pan2021electric,long2022toward}.

Quantum Byzantine agreement (QBA) is a promising approach for consensus problems. For three-party consensus, the first quantum solution using a three-qutrit singlet state was proposed in 2001~\cite{Fitzi2001quantum} and was experimentally demonstrated using a four-photon polarization-entangled state in 2008~\cite{gaertner2008xperimental}. This protocol and its subsequent protocols~\cite{fitzi2002detectable,Iblisdir2004byzantine,Neigovzen2008Multipartite,Rahaman2015Quantum,smania2016experimental} using some special entanglement, called detectable QBA framework, unavoidably weaken the two original Byzantine conditions with extra assumptions, which leads to a certain probability of aborting the protocol. More seriously, these rudimentary solutions are restricted to the three-party scenario and can only reach a one-bit message consensus~\cite{Fitzi2001quantum,gaertner2008xperimental,Iblisdir2004byzantine,Neigovzen2008Multipartite,Rahaman2015Quantum,smania2016experimental}. Some achievements have been made toward scalable multiparty QBA~\cite{Ben2005fast,taherkhani2018resource,sun2020multi,wang2022quantum} but their fault tolerance is $1/3$. In addition, QBA protocols require sophisticated techniques, such as multiparticle entanglement generation and distribution and entanglement swapping, which are difficult for practical implementations. Furthermore, the security of detectable QBA has not been proven rigorously~\cite{Gao:2008:Common,Gaertner:2008:Reply}.

\begin{table}[b]
	\caption{Comparison between \textcolor{black}{our work and detectable QBA framework.} \textcolor{black}{D-QBA: Detectable QBA.} N/A: not applicable. IC$_1$ $\&$ IC$_2$: two interactive consistency Byzantine conditions.}
	\resizebox{\linewidth}{!}{ 
		\begin{tabular}{ccc}
			\hline
			\hline
			\textbf{Performance}                                                                          & \textbf{This work}                                                               &\textbf{D-QBA~\cite{Fitzi2001quantum,gaertner2008xperimental,fitzi2002detectable,Iblisdir2004byzantine,Neigovzen2008Multipartite,Rahaman2015Quantum,smania2016experimental}}                                      \\ \hline
			Security analysis                            &  Yes      & N/A                    \\
			Fault tolerance                          & $N\ge2f+1, \forall f \in \mathbb{N}^+$  &  $N=3, f=1$      \\
			Message                   & Multiple       & Binary                 \\
			Decentralization                                                                                               & Yes                                                                              & N/A                            \\
			Entanglement                     & No                                                                  &Yes \\
			Strictly obey IC$_1$ \& IC$_2$                              & Yes                                                                              & No                       \\ \hline\hline
	\end{tabular}}
	\label{comparison}
\end{table}

Intriguingly, quantum entanglement is not necessary to break the $1/3$ fault-tolerance bound and  weaker correlation can also work. Here, different from detectable QBA, we propose a strict information-theoretical secure Byzantine agreement framework that exploits the recursion structure~\cite{kleinberg2006algorithm} and quantum digital signatures (QDS)~\cite{gottesman2001quantum,dunjko2014quantum,roehsner2018quantum} to address the limitation of  fault-tolerance bound and security loopholes (see Table~\ref{comparison}). It completely breaks the $1/3$ fault-tolerance bound with a fault tolerance of $N\ge2f+1, \forall f\in \mathbb{N}^+$ while strictly obeying IC$_1$ and IC$_2$  due to multiparty correlation provided by QDS. Our work is highly adaptable, because it can be achieved by any type of QDS, including the original proposal of  GC01-QDS~\cite{gottesman2001quantum} and its variants with \textcolor{black}{such as orthogonal encoding~\cite{amiri2016secure,Puthoor2016Mea,roberts2017experimental,collins2017experimental,an2019practical,thornton2019continuous,richter2021agile,qin2022quantum,yin2017experimental} and non-orthogonal encoding~\cite{yin2016practical,yin2017experiment,lu2021efficient,Weng2021secure}, and OTUH-type QDS~\cite{yin2021experimental,li2023one}. As QDS advances by leaps and bounds, it only requires coherent states instead of complex multiparticle entanglement and quantum memory~\cite{dunjko2014quantum,yin2016practical,amiri2016secure}.} Generating and maintaining entanglement is a sticking point in experimental setups, and the ability to relax this requirement can reduce the complexity of consensus systems and serve as a foundation for further research. Furthermore, our protocol is able to achieve consensus on multiple messages. In addition, we implement proof-of-principle experiments of the three-party and five-party consensus with three different QDS protocols, \textcolor{black}{BB84 GC01 QDS~\cite{amiri2016secure}, OTUH-QDS~\cite{yin2021experimental} and OTUH-QDS without perfect keys~\cite{li2023one}. }

\begin{figure}[t]
	\centering
	\includegraphics[width=0.45\textwidth]{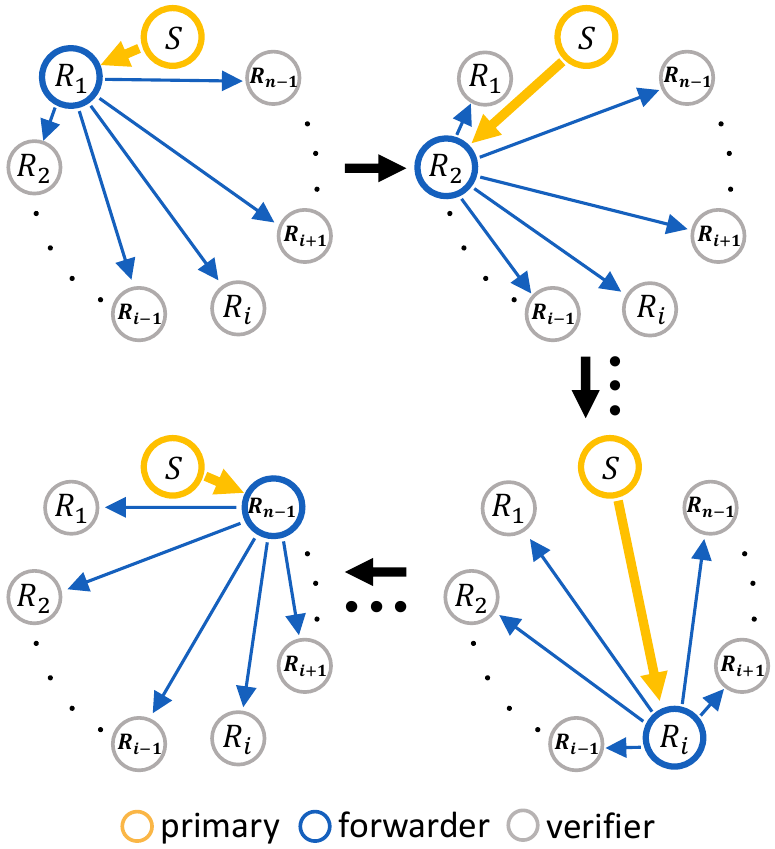}
	\caption{\textbf{Schematic of a multicast round including $n$ players with primary $S$.} The primary $S$ signs and then multicasts his message to the backups. The forwarder is chosen from among the backups, and unchosen backups act as verifiers. The primary, the forwarder and one of the verifiers perform a three-party QDS. The backups take turns acting as the forwarder. The arrow indicates the direction of the message delivery.}
	\label{fig2}
\end{figure}

\section{Results}
\subsection{Protocol definition}
Before stating our QBA framework, we introduce the \emph{multicast round}. In a multicast round with $n$ players, there is a primary, and the others are backups. The primary multicasts his or her message to the backups by the following operation, as shown in Fig~\ref{fig2}. One of the backups is selected as the forwarder, and the other unchosen backups become verifiers. The primary, the forwarder and one of the verifiers perform a three-party QDS to transmit the message. QDS is divided into two stages---distribution and messaging stage. The distribution stage is to distribute correlated quantum keys to the players. The messaging stage uses some classical operations and quantum keys to complete digital signatures. The messaging stage contains three steps: signing, forwarding, and verification. The primary signs the message, then sends the message and corresponding signature to the forwarder. After that, the forwarder will forward the message and signature to the verifier. Only when both the forwarder and verifier accept the signature, the signing is successful, i.e., the primary cannot deny the fact that she signed the message (nonrepudiation), and the message cannot be forged by others including the forwarder (unforgeability) (See Materials and Methods for details). For a chosen forwarder, the verifiers take turns participating in such three-party QDS. 
The above process will be repeated until all backups have acted as the forwarder one time. In the end, each backup records a list of $n-1$ messages, consisting of one message directly from the primary and $n-2$ messages forwarded by other backups. We call this list of messages broadcasting list in the later. Note that a complete multicast round  consists of three steps: ($\romannumeral1$) sign and multicast, ($\romannumeral2$) forward and ($\romannumeral3$) verify and record.

\begin{table}
	\caption{Protocol Definition.}
	
	\begin{tabular}{l}
		\rule[0.1em]{8.5cm}{0.5pt}
	\end{tabular}
	\textbf{Broadcasting phase}  
	\begin{flushleft}
		The broadcasting phase begins with $d=1$ and ends with $d=f$. We consider a general case $\mathcal{MR}^d_{\zeta}$ ($d$=$1,2,3,\cdots,f$):
	\end{flushleft}
	\begin{description}
		\item[1. Sign and Multicast]
		The primary signs and then multicasts the message $m^d_{\zeta}$ to the $N-d$ backups via QDS as shown in Fig.~\ref{fig2}.
		
		\item[2. Consistency check]
		If $d$$=$$1$, \textcolor{black}{there is no consistency check and the players will skip to \textit{Step 3}.} If $1<d\le f$, upon receiving the message from the primary, forwarder $R_j$ checks the consistency between it and the message that he received from the primary at the previous depth $d-1$. ($R_j$ visits the players who do not appear in route $\zeta$.) If consistency check is passed, perform \textit{Step 3}. Otherwise, he requests that this primary perform \textit{Step 1} again until he receives a consistent message. 
		\item[3. Forward]
		$R_j$ forwards the message to verifiers $R_k$ ($R_k$ visit the players except $R_j$ and those who have appeared in route $\zeta$).
		
		\item[4. Verify and record]
		Forwarder $R_j$ and verifier $R_k$ verify the message and corresponding signature. When both of them accept, the signature is successful and they add this valid message $m^d_{\zeta}$ to their own broadcasting lists $B^{d,R_j}_{\zeta}$ and $B^{d,R_k}_{\zeta}$, respectively. 
		
		\item[5. Recursion]
		
		The forwarder $R_j$ acts as the primary of $\mathcal{MR}^{d+1}_{\zeta \to R_j}$, and then repeat the above four steps. The recursion process ends up when $d=f$. 
	\end{description}
	
	\textbf{Gathering phase}
	
	\begin{flushleft}
		For the lieutenants $R_i$ ($i=1,2,\cdots,N-1$):
	\end{flushleft}
	\begin{description}
		\item[1. Input]
		In the bottom layer $d=f$, $R_i$ obtains the initial gathering lists $G^{f,R_i}_\zeta$$=$$B^{f,R_i}_\zeta$.
		
		\item[2. Recursion]
		When $1$$\le$$d$$<$$f$, the gathering lists at the corresponding depth and route are $G^{d,R_i}_{\zeta}$$=$$\bigcup_{R_p} \{m^{d+1,R_i}_{\zeta \to R_p}\}$, where $m^{d+1,R_i}_{\zeta \to R_p}$$=$${\rm majority}(G^{d+1,R_i}_{\zeta \to R_p})$ and $R_p$ visits all players except those who have appeared in route $\zeta$.
		
		\item[3. Output]
		$m_{S}^{1,R_i}={\rm majority}(G^{1,R_i}_{S})$.
	\end{description}
	
	\begin{tabular}{l}
		\rule[0.1em]{8.5cm}{0.5pt}
	\end{tabular}
	\label{table_protocol}
\end{table}

Generally, our QBA framework consists of two phases, namely, \emph{broadcasting phase} and \emph{gathering phase}. Suppose there is a system of total $N$ players including $f$ malicious ones. The commanding general (initial primary) is denoted as $S$, and the lieutenants are denoted as $R_i$, for $i=1,2,\cdots,N-1$. The flow chart of the two phases are shown in Table~\ref{table_protocol}. The broadcasting phase is designed for $R_i$ to exchange the message received from $S$ with each other, and the gathering phase is designed for $R_i$  to deduce the original message of $S$ according to the information gathered by themselves.

\emph{Broadcasting phase.} The broadcasting phase consists of successive multicast rounds. For clarity, we denote the multicast round as $\mathcal{MR}^{d}_{\zeta}$, where $\zeta$ represents the route of delivering the message and $d$ is the depth of the multicast round. The first multicast round started by the commanding general $S$ is denoted as $\mathcal{MR}^{1}_{S}$. In $\mathcal{MR}^{1}_{S}$, $S$ signs and then multicasts his message $m^1_S$ to all the lieutenants $R_i$. In the multicast round of next depth, $\mathcal{MR}^{2}_{S \to R_i}$, $R_i$ acts as a primary, and then signs and multicasts the message $m^{2}_{S \to R_i}$, which is he received from $S$, to the other lieutenants. The process will be repeated until $d=f$. We denote a list for the lieutenants $R_{i}$ to record the messages received by him in $\mathcal{MR}^d_{\zeta}$ as aforementioned broadcasting list $B^{d,R_i}_\zeta$. 

In the broadcasting phase, the consistency check occurs between Step \textit{1} and \textit{3}. Consider a general case: in $\mathcal{MR}^d_{\zeta}$, the primary signs and multicasts the message $m^d_{\zeta}$ to the backups, assuming that $R_{j}$ acts as the forwarder and $R_k$ acts as a verifier. In the multicast round at next depth $d+1$, $\mathcal{MR}^{d+1}_{\zeta\to R_j}$, $R_j$ will act as a primary and $R_k$ will act as the forwarder. The messages $R_j$ delivers to $R_k$ in the two rounds, $\mathcal{MR}^d_{\zeta}$ and $\mathcal{MR}^{d+1}_{\zeta\to R_j}$, must be consistent, because $R_k$ can check the consistency of the two messages. If the two messages are inconsistent, $R_k$  will reject them and ask $R_{j}$ to repeat the process until the two messages are consistent.

\emph{Gathering phase.} The deterministic function we used in the gathering phase is called the majority function. It outputs the value of the majority element in the input set (see Materials and Methods). In $\mathcal{MR}^d_\zeta$, the gathering list held by the lieutenant $R_i$, denoted as $G^{d,R_i}_\zeta$, is used for $R_i$ to deduce the message delivered by the primary of $\mathcal{MR}^d_\zeta$. In the bottom layer $d=f$, $R_i$ directly sets his or her own gathering list to $G^{f,R_i}_\zeta$$=$$B^{f,R_i}_\zeta$ and outputs $m^{f,R_i}_\zeta$$=$${\rm majority}(G^{f,R_i}_\zeta)$. Then, $m^{f,R_i}_\zeta$ becomes an element of the gathering list of $d=f-1$. Considering general case where $1 \le d< f$, all elements of $R_i$'s gathering list $G^{d,R_i}_\zeta$ are deduced from the lists $G^{d+1,R_i}_{\zeta \to R_p}$ in multicast round $\mathcal{MR}^{d+1}_{\zeta \to R_p}$ ($R_p$ visits all players who do not appear in route $\zeta$). When $p$=$i$, this element is directly set as the message that $R_i$ received from the primary in $\mathcal{MR}^{d}_{\zeta}$. With the recursive process, the gathering phase ends up when $d=1$, and then $R_i$ outputs $m^{1,R_i}_S$$=$${\rm majority}(G^{1,R_i}_{S})$ as the final decision. Note that the broadcasting lists record the messages that are the lieutenants themselves actually received during the broadcast phase, and the gathering lists record the messages that are the lieutenants deduced according to the information of the previous depth. Only when $d=f$, the gathering lists are the same as the broadcasting lists, i.e., $G^{f,R_i}_\zeta$$=$$B^{f,R_i}_\zeta$.

\subsection{Experimental implementation.}
We show proof-of-principle experimental implementation of our QBA framework for reaching consensus on a decentralized digital ledger, one of the most important application of blockchain. The digital ledger is a 1.10 MByte document that is a virtual transaction including time, clients, merchants, commodity and the amount. It is converted into a binary string of bits. We denote the correct message as $m1$, and the incorrect messages as $m2$, $m3$, and so on.  

\begin{figure*}[t]
	\centering
	\includegraphics[width=1\textwidth]{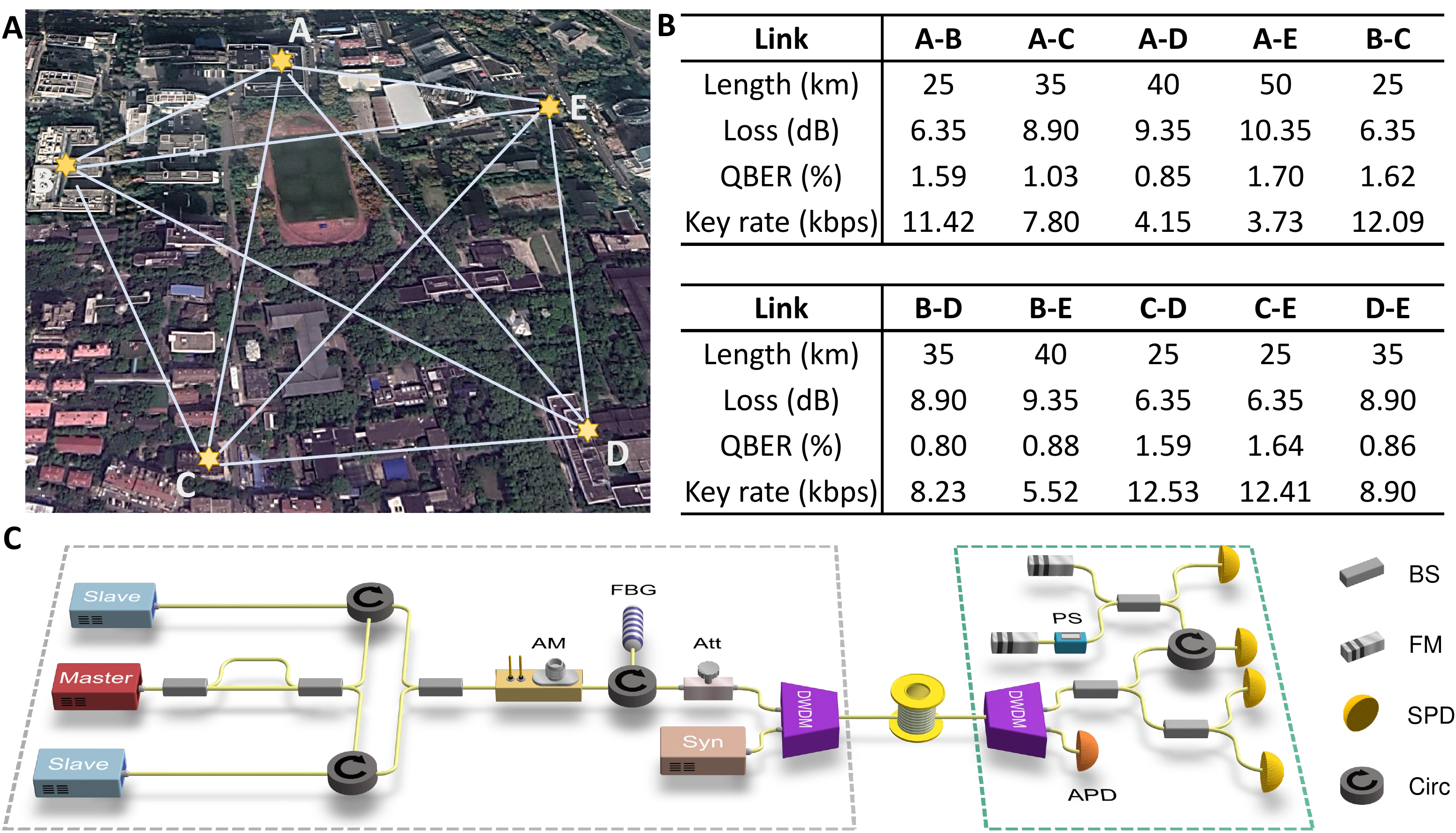}
	\caption{\textbf{Experimental implementation.}  \textbf{A.} The five players bring their own pre-distributed correlated quantum keys to five different buildings, and then perform the classical operations of the messaging stage. A-$S$, B-$R_1$, C-$R_2$, D-$R_4$ and E-$R_5$. In the messaging stage, the messages and corresponding signatures can be transmitted via authenticated classical channel. \textbf{B.} Main parameters of BB84 key generation in the laboratory. QBER: quantum bit error rate. \textbf{C.} Experimental setup of the four-intensity decoy-state quantum key generation system with a time-phase encoding. We take node A and B as an example. A uses a master laser, two slave lasers and an asymmetric interferometer to prepare optical pulses in the Z and X bases. An intensity modulator is used for the decoy-state modulating. Before passing through a set of filters, a monitor and an attenuator are utilized to regulate the photon number per pulse. B uses a biased beam splitter for the passive basis detection. The pulses either go directly to the time detector or pass through an asymmetric interferometer. A synchronization signal is distributed from node A to B through a wavelength division multiplexed quantum channel. BS: beam splitter; Circ: circulator; IM: intensity modulator; FBG: fibre Bragg grating; Att: attenuator; DWDM: dense wavelength division multiplexer; FM: Faraday mirror; PS: phase shifter; SPD: single-photon detector.}
	\label{map}
\end{figure*}

To show a high degree of adaptability of our work, we implement three-party consensus with single-bit GC01-QDS~\cite{amiri2016secure},  one-time universal$_2$ hashing (OTUH) QDS~\cite{yin2021experimental} and OTUH-QDS without perfect keys ~\cite{li2023one}, respectively (See Appendix~\ref{appB} for details). In addition, we utilize OTUH-QDS to realize the five-party consensus.  The key idea of  single-bit GC01-QDS is to first generate two pairs of raw quantum keys and then exchange half of the Bob's and Charlie's quantum keys with each other, which is called symmetrization step, to construct the correlation. On the contrary,  OTUH-QDS is to first construct the three-party correlation $X_a=X_b\oplus X_c $ and $Y_a=Y_b\oplus Y_c $ among the quantum keys of Alice (signer), Bob (forwarder) and Charlie (verifier). Alice signs the message with $X_a$ and $Y_a$, and then Bob and Charlie exchange their keys to complete the verification. These correlated raw quantum keys can be achieved by any quantum key generation process ~\cite{yin2020experimental,xu2020secure,pirandola2020advances,liu2021homodyne,lo2012measurement,lucamarini2018overcoming,xie2022breaking,fu2015long,cao2023realization,shen2023experimental}. Here, we utilize four-intensity decoy-state BB84 key generation process for the three QDS protocols~\cite{yin2020experimental}.

There are five independent players $S$ and $R_i$ ($i=1,2,3,4$). The correlated quantum keys of different pairwise players are pre-distributed in the laboratory via fiber spool, i.e., the distribution stage of QDS is completed in the laboratory.  They do not disclose any information of the their own quantum keys to others, and then bring the keys to five different buildings in Fig.~\ref{map} to simulate the real-life situation where the users are geographically separated, and then use these quantum keys to complete digital signatures, i.e.,  classical operations of the messaging stage are performed in the real locations. Note that, to simplify the proof-of-principle experiment, we employ the above method due to the immaturity of real-life multi-node quantum networks. It is anticipated that as quantum networks progress in maturity, enabling their widespread deployment and utilization, our QBA framework can be seamlessly integrated into practical quantum networks without the necessity of laboratory-based quantum key preparation.

\begin{figure*}[t]
	\centering
	\includegraphics[width=0.95\textwidth]{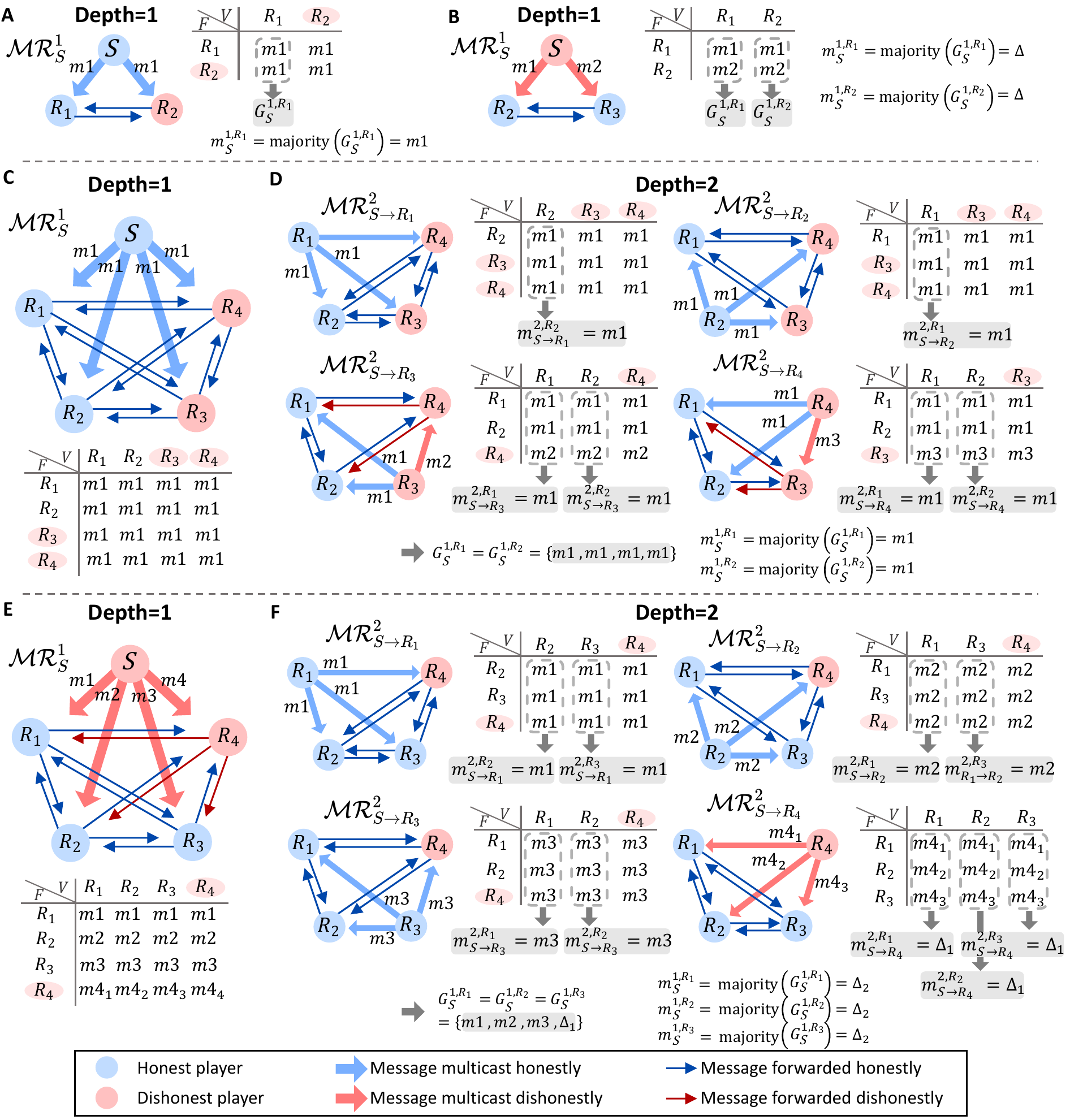}
	\caption{ \textbf{Experimental results for three-party and five-party consensus.} We use `F' to represent `forwarder' and `V' to represent `verifier' in the tables of lists. Each column of a table is a broadcasting list for the corresponding player. In the bottom layer $d=f$, $R_i$ sets his or her gathering list as $G^{f,R_i}_\zeta$$=$$B^{f,R_i}_\zeta$ and performs the gathering phase to deduce the final output. \textbf{A.} The multicast rounds of $d=1$ in three-party consensus with an honest primary. \textbf{B.} The multicast rounds at $d=1$ in three-party consensus with a dishonest primary. $\Delta$$=$${\rm majority}(m1,m2)$. \textbf{C(D).} The multicast rounds at $d=1$ ($d=2$) in five-party consensus with an honest initial primary.  \textbf{E(F).} The multicast rounds at $d=1$ ($d=2$) in five-party consensus with a dishonest initial primary. $\Delta_1$$=$${\rm  majority}(m4_1,m4_2,m4_3)$ and $\Delta_2$$=$${\rm majority}(m1,m2,m3,\Delta_1)$.}
	\label{results}
\end{figure*}

\begin{table}[b]
	\centering
	\caption{Consensus rates of our QBA framework adopting different QDS in the three-party consensus. The agreement rate, $CR$, is defined as the number of times a system can reach consensus per second. It can be expressed as $CR=\frac{SR}{C}$, where $C$ is communication complexity of the system, and $SR$ is the signature rate of adopted QDS. (See Materials and Methods).}
	\begin{tabular}{cc}
		\hline
		\hline
		\textbf{Different kinds of QDS}            & \textbf{Consensus rate}  \\\hline                                                         GC01-QDS~\cite{amiri2016secure}&         $4.5\times 10^{-8}$ \\
		OTUH-QDS~\cite{yin2021experimental}&         $11.95$ \\ 
		OTUH-QDS without perfect keys~\cite{li2023one}&          $6.12$\\    \hline\hline
	\end{tabular}
	\label{agreementrate}
\end{table}

$S$, $R_1$ and $R_2$ perform three-party consensus and all the five players perform five-party consensus. According to IC$_1$ and IC$_2$, we consider whether $S$ is honest or not. Here, we exemplify the three-party amd five-party consensus in Fig.~\ref{results}. Moreover, we show the consensus rates of our QBA framework when adopting these three QDS in Table~\ref{agreementrate}. OTUH-QDS, which can sign a multi-bit message each time, leads to much higher efficiency of QBA framework that the system can reach consensus 11.95 times per second, while single-bit GC01-QDS only reaches $4.5\times 10^{-8}$ times consensus per second under the same security parameter (See Appendix~\ref{appD} for calculation details).

(a) \textbf{The commanding general $S$ is honest in three-party consensus.} There is only one malicious player $R_2$ and thus only one layer $d=1$ in the three-party consensus. In $\mathcal{MR}^{1}_{S}$, $S$ sends correct message $m1$ via multicasting. $R_1$ records $m1$ when he acts as a forwarder, and records $m1$ received from $R_2$ when he acts as a verifier. The malicious $R_2$ must honestly forward $m1$ when he acts as a forwarder due to the unforgeability of QDS. Hence, as shown in Fig.~\ref{results}A, the gathering list of honest $R_1$, which is also the broadcasting list, is $G^{1,R_1}_{S}=B^{1,R_1}_{S}=\{m1,m1\}$. The final output of $R_1$ is $m^{1,R_1}_{S}={\rm majority}(G^{1,R_1}_{S})=m1$, which is consistent with the message sent by honest $S$. That satisfies IC$_1$.

(b) \textbf{The commanding general $S$ is dishonest in three-party consensus.} There is only one malicious player $S$. In $\mathcal{MR}^{1}_{S}$, $S$ sends conflicting messages $m1$ and $m2$ to honest $R_1$ and $R_2$, respectively. $R_1$ records $m1$ when he acts as a forwarder, and records $m2$ received from $R_2$ when he acts as a verifier. $R_2$ records $m2$ when he acts as a forwarder, and records $m1$ received from $R_1$ when he acts as a verifier. Hence, as shown in Fig.~\ref{results}B, the gathering list of honest $R_1$, which is also the broadcasting list, is $G^{1,R_1}_{S}=B^{1,R_1}_{S}=\{m1,m2\}$. The gathering list of honest $R_2$ is $G^{1,R_2}_{S}=B^{1,R_2}_{S}=\{m1,m2\}$, which is the same as that of $R_1$.  The final outputs of $R_1$ and $R_2$ are both $m^{1,R_1}_{S}=m^{1,R_2}_{S}={\rm majority}(\{m1,m2\})=\Delta$. Although the dishonest primary $S$ sends conflicting messages, honest $R_1$ and $R_2$ obtain the same output $\Delta$. That satisfies IC$_2$.

(c) \textbf{The commanding general $S$ is honest in five-party consensus.} There are two malicious players $R_3$ and $R_4$ ($f=2$), and there are two layers $d=1$ and $d=2$ in five-party consensus.

\textit{Broadcasting phase.} At depth $d=1$, in $\mathcal{MR}^{1}_{S}$, $S$ broadcasts the correct message $m_1$ via multicast process. In $\mathcal{MR}^{1}_{S}$, every one must forward the correct message $m_1$ to honest players because $S$ is honest and there are two honest players in QDS. Therefore, the broadcasting list of honest $R_1$ and $R_2$ are both $B^{1,R_1}_{S}=B^{1,R_2}_{S}=\{m1,m1,m1,m1\}$ (see Fig.~\ref{results}C). 

The results of the multicast processes at depth $d=2$ are shown in Fig.~\ref{results}D. In $\mathcal{MR}^{2}_{S \to R_1}$, because of the honest primary $R_1$ and Lemma 1 (see Materials and Methods for two lemmas), every one must forward the correct message $m_1$, and the broadcasting list of honest $R_2$ is $B^{2,R_2}_{S\to R_1}=\{m1,m1,m1\}$.  In $\mathcal{MR}^{2}_{S \to R_2}$, because of the honest primary $R_2$ and Lemma 1, every one must forward the correct message $m1$, and the broadcasting list of honest $R_1$ is $B^{2,R_1}_{S\to R_2}=\{m1,m1,m1\}$.  In $\mathcal{MR}^{2}_{S \to R_3}$, because of the dishonest primary $R_3$ and Lemma 2, only when dishonest $R_3$ and $R_4$ collude together, can $R_4$ successfully forward the conflicting message $m2$ to honest players. The broadcasting list of honest $R_1$ and $R_2$ are both $B^{2,R_1}_{S\to R_3}=B^{2,R_2}_{S\to R_3}=\{m1,m1,m2\}$. Similar to the analysis of $\mathcal{MR}^{2}_{S \to R_3}$, only when dishonest $R_3$ and $R_4$ collude together, $R_4$ can successfully forward the conflicting message $m_2$ to honest players. The broadcasting list of honest $R_1$ and $R_2$ are both $B^{2,R_1}_{S\to R_3}=B^{2,R_2}_{S\to R_3}=\{m1,m1,m3\}$.

\textit{Gathering phase.} At the bottom depth $d=2$, each player obtains the initial gathering lists  $G^{2,R_i}_\zeta=B^{2,R_i}_\zeta$. 

For $R_1$ at depth $d=2$, 
\begin{equation}
	\begin{aligned}
		\renewcommand{\arraystretch}{1.5}
		\begin{array}{ll}
			G^{2,R_1}_{S\to R_2}&=B^{2,R_1}_{S\to R_2}=\{m1,m1,m1\},\\
			G^{2,R_1}_{S\to R_3}&=B^{2,R_1}_{S\to R_3}=\{m1,m1,m2\},\\
			G^{2,R_1}_{S\to R_4}&=B^{2,R_1}_{S\to R_4}=\{m1,m1,m3\},\\
			m^{2,R_1}_{S\to R_1}& = m1,\\
			m^{2,R_1}_{S\to R_2}& = {\rm majority}(G^{2,R_1}_{S\to R_2})= m1,\\
			m^{2,R_1}_{S\to R_3}& = {\rm majority}(G^{2,R_1}_{S\to R_3})= m1,\\
			m^{2,R_1}_{S\to R_4}& = {\rm majority}(G^{2,R_1}_{S\to R_4})= m1,
		\end{array}	
	\end{aligned}
\end{equation}
where the message $m^{2,R_1}_{S\to R_1}$ is directly received from $S$ when $R_1$ acts as the forwarder. $R_1$ deduced that $m^{2,R_1}_{S\to R_2}$ is the message that $R_2$ received from $S$,  $m^{2,R_1}_{S\to R_3}$ is the message that $R_3$ received from $S$, and $m^{2,R_1}_{S\to R_4}$ is the message that $R_4$ received from $S$. Then, $m^{2,R_1}_{S\to R_1}$, $m^{2,R_1}_{S\to R_2}$, $m^{2,R_1}_{S\to R_3}$ and $m^{2,R_1}_{S\to R_4}$ constitute the gathering list of $R_1$ at depth $d=1$, where  $G^{1,R_1}_{S}=\{m^{2,R_1}_{S\to R_1},m^{2,R_1}_{S\to R_2},m^{2,R_1}_{S\to R_3},m^{2,R_1}_{S\to R_4}\}=\{m1,m1,m1,m1\}$. Thus, at depth $d=1$, $R_1$ obtains the final output 
\begin{equation}
	\begin{aligned}
		\renewcommand{\arraystretch}{1.2}
		\begin{array}{ll}
			m^{1,R_1}_{S}&={\rm majority}(G^{1,R_1}_{S}) \\
			&={\rm majority}(\{m1,m1,m1,m1\}) \\
			&=m1.
		\end{array}	
	\end{aligned}
\end{equation}

For $R_2$ at depth $d=2$, 
\begin{equation}
	\begin{aligned}
		\renewcommand{\arraystretch}{1.5}
		\begin{array}{ll}
			G^{2,R_2}_{S\to R_1}&=B^{2,R_2}_{S\to R_1}=\{m1,m1,m1\},\\
			G^{2,R_2}_{S\to R_3}&=B^{2,R_2}_{S\to R_3}=\{m1,m1,m2\},\\
			G^{2,R_2}_{S\to R_4}&=B^{2,R_2}_{S\to R_4}=\{m1,m1,m3\},\\
			m^{2,R_2}_{S\to R_1}& = {\rm majority}(G^{2,R_2}_{S\to R_1})= m1,\\
			m^{2,R_2}_{S\to R_2}& = m1,\\
			m^{2,R_2}_{S\to R_3}& = {\rm majority}(G^{2,R_2}_{S\to R_3})= m1,\\
			m^{2,R_2}_{S\to R_4}& = {\rm majority}(G^{2,R_2}_{S\to R_4})= m1,
		\end{array}	
	\end{aligned}
\end{equation}
where the message $m^{2,R_2}_{S\to R_2}$ is directly received from $S$ when $R_2$ acts as the forwarder. $R_2$ deduced that $m^{2,R_2}_{S\to R_1}$ is the message that $R_1$ received from $S$,  $m^{2,R_2}_{S\to R_3}$ is the message that $R_3$ received from $S$, and $m^{2,R_2}_{S\to R_4}$ is the message that $R_4$ received from $S$. Then, $m^{2,R_2}_{S\to R_1}$, $m^{2,R_2}_{S\to R_2}$, $m^{2,R_2}_{S\to R_3}$ and $m^{2,R_2}_{S\to R_4}$ constitute the gathering list of $R_2$ at depth $d=1$, where  $G^{1,R_2}_{S}=\{m^{2,R_2}_{S\to R_1},m^{2,R_2}_{S\to R_2},m^{2,R_2}_{S\to R_3},m^{2,R_2}_{S\to R_4}\}=\{m1,m1,m1,m1\}$. Thus, at depth $d=1$, $R_2$ obtains the final output 
\begin{equation}
	\begin{aligned}
		\renewcommand{\arraystretch}{1.2}
		\begin{array}{ll}
			m^{1,R_2}_{S}&={\rm majority}(G^{1,R_2}_{S}) \\
			&={\rm majority}(\{m1,m1,m1,m1\}) \\
			&=m1.
		\end{array}	
	\end{aligned}
\end{equation}

The final outputs of $R_1$ and $R_2$ are both $m^{1,R_1}_{S}=m^{1,R_2}_{S}={\rm majority}(\{m_1,m_1,m_1,m_1\})=m_1$, which is consistent with the initial message sent by the honest primary $S$. This result satisfies IC$_1$.  

(d) \textbf{The commanding general $S$ is dishonest in five-party consensus.} There are two malicious players $S$ and $R_4$ ($f=2$), and there are two layers $d=1$ and $d=2$ in five-party consensus.

\textit{Broadcasting phase.} At depth $d=1$, in $\mathcal{MR}^{1}_{S}$, the dishonest primary $S$ can broadcast the different messages $m1$, $m2$ and $m3$ to honest $R_1$, $R_2$ and $R_3$, respectively, via multicast process. Additionally, due to Lemma 2, malicious $S$ and $R_4$ can collude together and then $R_4$ can deliberately forward different messages $m4_1$, $m4_2$ and $m4_3$ to honest $R_1$, $R_2$ and $R_3$, respectively. Therefore, the broadcasting list of honest $R_1$, $R_2$ and $R_3$ are $B^{1,R_1}_{S}=\{m1,m2,m3,m4_1\}$, $B^{1,R_2}_{S}=\{m1,m2,m3,m4_2\}$ and $=B^{1,R_3}_{S}=\{m1,m2,m3,m4_3\}$, respectively (see Fig.~\ref{results}E). 

The results of the multicast processes at depth $d=2$ are shown in Fig.~\ref{results}F. In $\mathcal{MR}^{2}_{S \to R_1}$, because of the honest primary $R_1$ and Lemma 1, every one must forward $R_1$'s message $m_1$, and the broadcasting list of honest $R_2$ and $R_3$ are both $B^{2,R_2}_{S\to R_1}=B^{2,R_3}_{S\to R_1}=\{m1,m1,m1\}$.  In $\mathcal{MR}^{2}_{S \to R_2}$, because of the honest primary $R_2$ and Lemma 1, every one must forward $R_1$'s message $m2$, and the broadcasting lists of honest $R_1$ and $R_3$ are both $B^{2,R_1}_{S\to R_2}=B^{2,R_3}_{S\to R_2}=\{m2,m2,m2\}$.  In $\mathcal{MR}^{2}_{S \to R_3}$, because of the honest primary $R_2$ and Lemma 1, every one must forward $R_3$'s message $m3$, and the broadcasting lists of honest $R_1$ and $R_2$ are both $B^{2,R_1}_{S\to R_3}=B^{2,R_2}_{S\to R_3}=\{m3,m3,m3\}$. In $\mathcal{MR}^{2}_{S \to R_4}$, because the dishonest primary $R_4$ can successfully broadcast the conflicting messages $m4_1$, $m4_2$ and $m4_3$ to honest players $R_1$, $R_2$ and $R_3$, respectively. The broadcasting list of honest $R_1$, $R_2$ and $R_3$ are all $B^{2,R_1}_{S\to R_4}=B^{2,R_2}_{S\to R_4}=B^{2,R_3}_{S\to R_4}=\{m4_1,m4_2,m4_3\}$. 

\textit{Gathering phase.} At the bottom depth $d=2$, each player obtains the initial gathering lists  $G^{2,R_i}_\zeta=B^{2,R_i}_\zeta$. 

For $R_1$ at depth $d=2$, 
\begin{equation}
	\begin{aligned}
		\renewcommand{\arraystretch}{1.5}
		\begin{array}{ll}
			G^{2,R_1}_{S\to R_2}&=B^{2,R_1}_{S\to R_2}=\{m2,m2,m2\},\\
			G^{2,R_1}_{S\to R_3}&=B^{2,R_1}_{S\to R_3}=\{m3,m3,m3\},\\
			G^{2,R_1}_{S\to R_4}&=B^{2,R_1}_{S\to R_4}=\{m4_1,m4_2,m4_3\},\\
			m^{2,R_1}_{S\to R_1}& = m1,\\
			m^{2,R_1}_{S\to R_2}& = {\rm majority}(G^{2,R_1}_{S\to R_2})= m2,\\
			m^{2,R_1}_{S\to R_3}& = {\rm majority}(G^{2,R_1}_{S\to R_3})= m3,\\
			m^{2,R_1}_{S\to R_4}& = {\rm majority}(G^{2,R_1}_{S\to R_4})= \Delta_1,
		\end{array}	
	\end{aligned}
\end{equation}
where the message $m^{2,R_1}_{S\to R_1}$ is directly received from $S$ when $R_1$ acts as the forwarder. $R_1$ deduced that $m^{2,R_1}_{S\to R_2}$ is the message that $R_2$ received from $S$,  $m^{2,R_1}_{S\to R_3}$ is the message that $R_3$ received from $S$, and $m^{2,R_1}_{S\to R_4}$ is the message that $R_4$ received from $S$. Then, $m^{2,R_1}_{S\to R_1}$, $m^{2,R_1}_{S\to R_2}$, $m^{2,R_1}_{S\to R_3}$ and $m^{2,R_1}_{S\to R_4}$ constitute the gathering list of $R_1$ at depth $d=1$, where  $G^{1,R_1}_{S}=\{m^{2,R_1}_{S\to R_1},m^{2,R_1}_{S\to R_2},m^{2,R_1}_{S\to R_3},m^{2,R_1}_{S\to R_4}\}=\{m1,m2,m3,\Delta_1\}$. Thus, at depth $d=1$, $R_1$ obtains the final output
\begin{equation}
	\begin{aligned}
		\renewcommand{\arraystretch}{1.2}
		\begin{array}{ll}
			m^{1,R_1}_{S}&={\rm majority}(G^{1,R_1}_{S}) \\
			&={\rm majority}(\{m1,m2,m3,\Delta_1\}) \\
			&=\Delta_2.
		\end{array}	
	\end{aligned}
\end{equation}

For $R_2$ at depth $d=2$, 
\begin{equation}
	\begin{aligned}
		\renewcommand{\arraystretch}{1.5}
		\begin{array}{ll}
			G^{2,R_2}_{S\to R_1}&=B^{2,R_2}_{S\to R_1}=\{m1,m1,m1\},\\
			G^{2,R_2}_{S\to R_3}&=B^{2,R_2}_{S\to R_3}=\{m3,m3,m3\},\\
			G^{2,R_2}_{S\to R_4}&=B^{2,R_2}_{S\to R_4}=\{m4_1,m4_2,m4_3\},\\
			m^{2,R_2}_{S\to R_1}& = {\rm majority}(G^{2,R_2}_{S\to R_2})= m1,\\
			m^{2,R_2}_{S\to R_2}& = m2,\\
			m^{2,R_2}_{S\to R_3}& = {\rm majority}(G^{2,R_2}_{S\to R_3})= m3,\\
			m^{2,R_2}_{S\to R_4}& = {\rm majority}(G^{2,R_2}_{S\to R_4})= \Delta_1,
		\end{array}	
	\end{aligned}
\end{equation}
where the message $m^{2,R_2}_{S\to R_2}$ is directly received from $S$ when $R_2$ acts as the forwarder. $R_2$ deduced that $m^{2,R_2}_{S\to R_1}$ is the message that $R_1$ received from $S$,  $m^{2,R_2}_{S\to R_3}$ is the message that $R_3$ received from $S$, and $m^{2,R_2}_{S\to R_4}$ is the message that $R_4$ received from $S$. Then, $m^{2,R_2}_{S\to R_1}$, $m^{2,R_2}_{S\to R_2}$, $m^{2,R_2}_{S\to R_3}$ and $m^{2,R_2}_{S\to R_4}$ constitute the gathering list of $R_2$ at depth $d=1$, where  $G^{1,R_2}_{S}=\{m^{2,R_2}_{S\to R_1},m^{2,R_2}_{S\to R_2},m^{2,R_2}_{S\to R_3},m^{2,R_2}_{S\to R_4}\}=\{m1,m2,m3,\Delta_1\}$. Thus, at depth $d=1$, $R_2$ obtains the final output
\begin{equation}
	\begin{aligned}
		\renewcommand{\arraystretch}{1.2}
		\begin{array}{ll}
			m^{1,R_2}_{S}&={\rm majority}(G^{1,R_2}_{S}) \\
			&={\rm majority}(\{m1,m2,m3,\Delta_1\}) \\
			&=\Delta_2.
		\end{array}	
	\end{aligned}
\end{equation}

For $R_3$ at depth $d=2$, 
\begin{equation}
	\begin{aligned}
		\renewcommand{\arraystretch}{1.5}
		\begin{array}{ll}
			G^{2,R_3}_{S\to R_1}&=B^{2,R_2}_{S\to R_1}=\{m1,m1,m1\},\\
			G^{2,R_3}_{S\to R_2}&=B^{2,R_2}_{S\to R_2}=\{m2,m2,m2\},\\
			G^{2,R_3}_{S\to R_4}&=B^{2,R_2}_{S\to R_4}=\{m4_1,m4_2,m4_3\},\\
			m^{2,R_3}_{S\to R_1}& = {\rm majority}(G^{2,R_3}_{S\to R_2})= m1,\\
			m^{2,R_3}_{S\to R_2}& = {\rm majority}(G^{2,R_3}_{S\to R_2})=m2,\\
			m^{2,R_3}_{S\to R_3}& = m3,\\
			m^{2,R_3}_{S\to R_4}& = {\rm majority}(G^{2,R_3}_{S\to R_4})= \Delta_1,
		\end{array}	
	\end{aligned}
\end{equation}
where the message $m^{2,R_2}_{S\to R_2}$ is directly received from $S$ when $R_3$ acts as the forwarder. $R_3$ deduced that $m^{2,R_3}_{S\to R_1}$ is the message that $R_1$ received from $S$,  $m^{2,R_3}_{S\to R_2}$ is the message that $R_2$ received from $S$, and $m^{2,R_3}_{S\to R_4}$ is the message that $R_4$ received from $S$. Then, $m^{2,R_3}_{S\to R_1}$, $m^{2,R_3}_{S\to R_2}$, $m^{2,R_3}_{S\to R_3}$ and $m^{2,R_3}_{S\to R_4}$ constitute the gathering list of $R_3$ at depth $d=1$, where  $G^{1,R_3}_{S}=\{m^{2,R_3}_{S\to R_1},m^{2,R_3}_{S\to R_2},m^{2,R_3}_{S\to R_3},m^{2,R_3}_{S\to R_4}\}=\{m1,m2,m3,\Delta_1\}$. Thus, at depth $d=1$, $R_3$ obtains the final output
\begin{equation}
	\begin{aligned}
		\renewcommand{\arraystretch}{1.2}
		\begin{array}{ll}
			m^{1,R_3}_{S}&={\rm majority}(G^{1,R_3}_{S}) \\
			&={\rm majority}(\{m1,m2,m3,\Delta_1\}) \\
			&=\Delta_2.
		\end{array}	
	\end{aligned}
\end{equation}

The final outputs of honest $R_1$, $R_2$ and $R_3$ are all $m^{1,R_1}_{S}$$=$$m^{1,R_2}_{S}$$=$$m^{1,R_3}_{S}$$=$$\Delta_2$. Although the dishonest primary $S$ sends conflicting messages, the honest players $R_1$, $R_2$ and $R_3$ obtain the same output $\Delta_2$. This result satisfies IC$_2$.

In conclusion, our experimental results show that in real three-party and five-party consensus, our protocol can not only satisfy the two original Byzantine conditions IC$_1$ and IC$_2$, but also achieve the fault tolerance of $n\ge 2f+1$ which breaks the 1/3 fault-tolerance lower bound.

\section{Discussion}
The $1/3$ fault-tolerance bound cannot be beaten for any arbitrary pairwise communication~\cite{pease1980reaching,dolev1986possibility,fischer1986easy,fitzi2001advances}; not even quantum channels can help solve this problem. If the nodes of a system are linked by the channels that are independent of each other, the bound is unable to be beaten. Intriguingly, when quantum entanglement is introduced into the system, it is possible to surpass this bound because quantum entanglement provides the correlation and removes the independence~\cite{Fitzi2001quantum}. Although detectable QBA framework is designed according to multi-particle entanglement~\cite{Fitzi2001quantum,gaertner2008xperimental,fitzi2002detectable,Iblisdir2004byzantine,Neigovzen2008Multipartite,Rahaman2015Quantum,smania2016experimental}, they cannot extend to more than three participants and unavoidably weaken the Byzantine conditions, because multi-particle entangled states are very hard to prepare and maintain, and these protocols do not fully utilize the correlation to protect the unforgeability and nonrepudiation which leads to a certain probability of failure.

Is quantum entanglement necessary or can weaker multiparty correlation work in Byzantine agreement? Fortunately, QDS is a useful tool for solving this problem due to its asymmetric relationship among three players. In addition, three-party QDS is naturally decentralized due to its structure without a fully trusted third party. The two essential properties of QDS, unforgeability and nonrepudiation, effectively curtail the malevolent activities of malicious players within the system, preventing them from deliberately delivering conflicting messages. Asymmetric relationship of QDS makes the channels no longer independent of each other. Consequently, our protocol can break the $1/3$ fault-tolerance bound. 

Note that the most important thing to break the fault-tolerance bound is to provide a decentralized multiparty correlation to remove the independence of pairwise channels. Intriguingly, quantum entanglement and asymmetric relationship of QDS both satisfy the above requirement. In addition, if we can find a three-party information-theoretically secure classical digital signature scheme which has the same decentralized structure as QDS~\cite{amiri2018efficient}, our framework can also break the bound. However, up to now, we do not find classical correlation greater than QDS without any additional assumptions because these classical schemes require extra assumptions such as the existence of a trusted third party and authenticated broadcast channels which disobeys decentralization of Byzantine agreement~\cite{wallden2015quantum}. By bridging two prominent research themes, the Byzantine agreement
and quantum digital signatures, our work paves the way for practical quantum blockchain and quantum consensus networks.

\textcolor{black}{In the end, we want to highlight that although our QBA framework utilizes quantum digital signature, it is still unable to surpass the famous blockchain trilemma (see Materials and Methods).The blockchain trilemma highlights the intricate balance required among three fundamental attributes: decentralization, security, and scalability. Within our QBA framework, we have achieved notable success in security, surpassing the  1/3 fault-tolerance bound while maintaining information-theoretical security. Moreover, our approach has fully decentralization. However, it falls short in scalability, exhibiting exponential communication complexity, as shown in Eq.~(\ref{complexity}). Consequently, our research remains bound by the constraints posed by the blockchain trilemma. Intriguing inquiries linger as we ponder the blockchain trilemma's resilience to quantum resources. Is it an irrefutable theorem or an assailable postulate? The possibility of quantum resources challenging the blockchain trilemma beckons us towards further scholarly exploration.}

\section{Materials and Methods}

\subsection{Quantum Digital Signatures}
Quantum digital signatures with information-theoretical security have two major properties, nonrepudiation and unforgeability. They are all divided into the two stages, distribution stage and messaging stage. The distribution stage is to distribute the correlated raw quantum keys of Alice-Bob and Alice-Charlie for the messaging stage. The correlated quantum keys can be achieved by some classical operations, such as symmetrization step used in CV-QDS~\cite{richter2021agile}, BB84 GC01-QDS~\cite{amiri2016secure}, and MDI-QDS~\cite{yin2017experimental,roberts2017experimental}, test bits used in SARG04-QDS~\cite{yin2016practical,lu2021efficient,Weng2021secure}, and secret sharing used in OTUH-QDS~\cite{yin2021experimental,li2023one}. The messaging stage is to complete the digital signature to determine whether it is successful or not.

The brief process of messaging stage can be described as follows. Alice is a `signer'. Bob is a `forwarder'. Charlie is a `verifier'.  Alice signs a message with her quantum keys, and then transmits the message and corresponding signature to Bob. Bob forwards the message and corresponding signature to the verifier Charlie. Then, Bob and Charlie will check the message and corresponding signature, respectively. The process of QDS is successful when and only when both Bob and Charlie accept the message and the corresponding signature. The signature rate, $SR$, is defined as the number of times the players can perform the QDS per second.

{\it{Nonrepudiation.}}
Nonrepudiation refers to a situation in which the signer cannot successfully dispute the authorship of his signature. This means that Alice cannot deny the fact that she signed the message if the signature is accepted by both Bob and Charlie.

{\it{Unforgeability.}}
Unforgeability refers to a situation in which no one can forge a message and its corresponding signature. This means that if Bob forwards a forged message and signature, it will be impossible for him to successfully make Charlie accept the forged message and signature.

\subsection{Majority Function}
The majority function we apply in our protocol aims to output the element that appears most often for an input set.  For example, when the input set is $M=\left \{ m_1, m_1,m_1, m_2, m_2\right \} $, the output will be ${\rm majority} (M)=m_1$. In a few cases, more than one element appears most frequently in the input set, and the systems that calculate the majority function on the input set are often deliberately biased toward one of them that we set initially. For example, when the input set is $M=\left \{ m_1, m_1,m_1, m_2, m_2, m_2\right \} $, the output will be ${\rm majority} (M)$$=$$m_1$ ($m_2$), which is determined by the biased output $m_1$ ($m_2$) that we set before calculation. Note that for the same input sets with different players, the majority function outputs the same value, which we will denote as $\Delta$.

\subsection{Communication Complexity and Consensus Rate}
To measure the consumed resources, we define the number of times the QDS process is implemented to reach consensus as the communication complexity, denoted as $C$. The total communication complexity of our QBA protocol can be expressed by
\begin{equation}
	\begin{aligned}
		C  = &\arrange{N-1}{2} + (N-1)\arrange{N-2}{2}+(N-1)(N-2)\arrange{N-3}{2}\\
		& +\cdots+(N-1)(N-2)\cdots(N-f+1)\arrange{N-f}{2}\\
		= &\sum_{m = 0}^{f-1}\arrange{N-1}{2+m},
	\end{aligned}
	\label{complexity}
\end{equation}
where $\arrange{a}{b}=\frac{a!}{(a-b)!}$ is $b$-permutations of $a$, $f$ is the number of dishonest players and $N$ is the number of all players. Here $\arrange{n-1}{2+m}$ represents the communication complexity at depth $m+1$. In reality, we need to perform $\left [ \frac{N-1}{2} \right ]$ recursions for a real-life consensus system with unknown $f$, where $\left [ x \right ]$ is the greatest integer less than or equal to $x$.

Consider the simple case that the system uses the QDS protocol which has the same signature rate, denoted as $SR$, in all the multicast rounds. We define the consensus rate of our QBA protocol as
\begin{equation}
	CR=\frac{SR}{C}=\frac{SR}{\sum_{m = 0}^{f-1}\arrange{N-1}{2+m}},
\end{equation}
where $C$ is the communication complexity of the system. $CR$ is the important index to indicate the efficiency of QBA. To get the higher consensus rate, we need to adopt the QDS protocol that has the higher signature rate. We can find that as the increase of the total number of players and the number of malicious players, the communication complexity will increase, which leads to the decrease of the consensus rate.

\subsection{Two Important Lemmas in Security Analysis}
In our protocol, the performance of honest (dishonest) players follows the same rule. Therefore, the players can be divided into two groups, the honest and dishonest. Also, the elements of a gathering list can be divided in the same way. Therefore, we can simplify the protocol with a perfect binary tree model where one tree node represents the set of multicast rounds with honest or dishonest primaries. The left (right) child tree node represents the multicast rounds with honest (dishonest) primaries of the next depth. In what follows, when we say a tree node is honest(dishonest), it means that the primaries in this tree node is honest (dishonest). And we can obtain the important Lemma 1 and Lemma 2. The proofs for two lemmas and complete security analysis for our protocol can be found in Appendix~\ref{appC} in detail.

\textit{Lemma 1}: Suppose that B is a right child tree node of a parent node A who is honest, and C is the left child tree node of B. The messages delivered in C are consistent with those of A, which protects the consistency of the delivered messages. 

\textit{Lemma 2}: Suppose that B is a right child tree node of a parent node A who is honest, and E is the right child tree node of B. The message multicast in E can be inconsistent with those of A, which disrupts the consistency of the delivered messages.

\subsection{Experimental Setting}
The master laser generates phase-randomized 1.6 ns-wide laser pulses with a repetition rate of 100 MHz at 1550.12 nm. The system frequency is 100 MHZ, but due to the 400 ns dead time every 10 us, the effective frequency of optical pulse is 96 MHz. Two pairs of pulses with relative phases 0 and $\pi$ at a 2 ns time delay generated by an asymmetric interferometer are injected into two slave lasers through the optical circulator, respectively. By controlling the trigger electrical signal of two slave lasers, Alice randomly prepares quantum states in the Z (time) and X (phase) bases by using 400 ps-wide slave laser pulses. The programmable delay chip with a 10 ps timing resolution is used to calibrate the time consistency. The spectral consistency is naturally satisfied because of the laser seeding technique~\cite{comandar2016quantum}. A 50 GHz nominal bandwidth fiber Bragg grating is used to remove extra spurious emission and precompensate for the pulse broadening in the fiber transmission. The 2 ns-wide synchronization pulses with repetition rates of 100 kHz are transmitted via the quantum channel using multiplexed wavelength division. The intensities are set as $\mu=0.40$, $\nu=0.20$, $\omega=0.4$ and $0$ with the corresponding probabilities $p_{\mu}=0.60$, $p_{\nu}=0.20$, $p_{\omega}=0.15$ and  $p_{0}=0.05$, respectively. If trigger signal is not provided to the slave laser, the vacuum state is generated . The amplitude modulator generates two different intensities, and the intensity of $\omega$ is double that of $\nu$ ($\omega=2\nu$) since it has two pulses in the X basis. At the receiving end, a 30:70 biased beam splitter is used to perform passive basis detection after a wavelength division demultiplexer. A probability of 30\% is measured in the phase basis and the probability of 70\% is used to receive in the time basis. A Faraday-Michelson interferometer is used for the phase measurement, in which phase drift is compensated in real time by using the phase shifter. The total insertion losses of the time and phase bases are 4.25 and 8 dB, respectively. The efficiency of single-photon detectors is 20\% at a 160 dark count per second. To decrease the after-pulse probability, we set the dead times to 10 $\mu$s for the links.

\subsection{Blockchain Trilemma}
\textcolor{black}{Blockchain possesses three crucial attributes: decentralization, security, and scalability. Decentralization forms the fundamental core of blockchain technology, emphasizing its inherent nature. Security stands as a paramount concern in any blockchain system, while scalability presents a formidable challenge. However, the blockchain trilemma emerges from the inherent difficulty of achieving a harmonious balance among these three essential elements. Blockchains are often forced to make trade-offs that prevent them from achieving all the three aspects. Note that the trilemma is just a model to conceptualize the various challenges facing blockchain technology. There is no strict proof that the 3 aspects cannot be achieved. But to date, there are no protocols able to break down the trilemma. The design needs to weaken the requirements for a certain feature.}

\section*{Acknowledgments}

\subsection*{General} 
We thank Y. Fu and M.-G. Zhou for their valuable discussions.

\subsection*{Author Contributions} 
H.-L.Y. and Z.-B.C. conceived the research. C.-X. W., R.-Q. G. and H.-L.Y. developed the quantum Byzantine agreement. R.-Q. G., C.-X. W. and H.-L.Y. provided the security proof. Y. B., Y.-S.L. and H.-L.Y. performed the quantum communication network. R.-Q. G., C.-X. W. and H.-L.Y. performed quantum consensus of digital ledger. C.-X. W., R.-Q. G. and H.-L.Y. co-wrote the manuscript, with input from the other authors. All authors have discussed the results and proofread the manuscript.

\subsection*{Funding}
This study was supported by the National Natural Science Foundation of China (No. 12274223), the Natural Science Foundation of Jiangsu Province (No. BK20211145), the Fundamental Research Funds for the Central Universities (No. 020414380182), the Key Research and Development Program of Nanjing Jiangbei New Area (No. ZDYD20210101), the Program for Innovative Talents and Entrepreneurs in Jiangsu (No. JSSCRC2021484).

\subsection*{Conflicts of Interest}
The authors declare that they have no competing interests.

\subsection*{Data Availability}
Data  generated  and  analyzed  during  the  current  study  are  available  from  the corresponding  author  upon  reasonable  request.

\appendix
\section{Some pre-knowledge of Byzantine agreement}\label{appA}
\subsection{Blockchain}
Blockchain is a decentralized digital database technology that allows secure transactions between multiple parties without the need for intermediaries. It was first introduced in 2008 as the underlying technology for the cryptocurrency, Bitcoin. However, its potential applications have expanded beyond just cryptocurrencies. What makes blockchain unique is that it is a distributed system, which means that it is maintained by a network of nodes that are interested in maintaining it rather than a central authority. Every participant in the network holds a copy of the blockchain, and any changes to the database require consensus among the nodes. This makes it virtually impossible for a single entity to control or manipulate the blockchain.

The potential applications of blockchain are vast and include everything from cryptocurrency, financial transactions (digital ledgers), the Internet of Things and supply chain management to digital identity verification and voting systems. Its decentralized and secure nature makes it an attractive solution for businesses and organizations looking to streamline processes, increase efficiency, and reduce costs. 

Blockchain includes many cryptography tasks, such as consensus, timestamp, identity authentication, privacy protection and so on. The most important one of them is the consensus problem, known as the Byzantine general problem, which is the research topic of our work. Our work does not aim to solve all the cryptography tasks of blockchain, and we focus on the core problem, Byzantine consensus.

\subsection{Byzantine general problem}
The Byzantine General Problem (also called Byzantine fault tolerance problem) is a classic computer science problem that deals with the challenge of coordinating a group of distributed and autonomous entities to reach a consensus in the presence of faulty or malicious actors~\cite{lamport1982byzantine}. In this problem, a group of Byzantine generals is camped outside a city and must coordinate their attack or retreat plans via messengers. However, some of the generals may be traitors who aim to sabotage the coordination, and messengers can be captured or corrupted during transmission, leading to false messages.

The challenge is to design a Byzantine agreement protocol that ensures that all loyal generals agree on a common plan of action, even in the presence of faulty or malicious actors. This problem has applications in distributed computing, cryptography, and especially blockchain technology. The Byzantine General Problem remains an active research topic in computer science and is considered a fundamental problem in distributed systems.

\subsection{Two necessary interactive consistency (IC) Byzantine condition}
Lamport et.al have proven that the Byzantine General Problem can be translated in a `commanding general-lieutenants' model, where the commanding general is randomly chosen from among all Byzantine generals and the others become lieutenants to reach consensus on the order of the commanding general~\cite{lamport1982byzantine}. A strict Byzantine agreement must satisfy the following two interactive consistency Byzantine conditions as follows. IC$_1$: All loyal lieutenants obey the same order. IC$_2$: Every loyal lieutenant obeys the order he or she sends if the commanding general is loyal. These two conditions emphasize two major concerns. When the commanding general is dishonest, all loyal players output consistent values. When the commanding general is honest, all loyal players output consistent and correct values. A strict Byzantine agreement must obey these two original conditions without adding any other assumptions. However, for detectable QBA protocols to achieve three-party consensus, an extra assumption is needed: there must be a certain probability that the protocol will fail. The players must discard the outcome when the protocol fails and perform the process again until the protocol succeeds. Therefore, all detectable QBA protocols are weaker versions of the Byzantine agreement.

\section{Quantum digital signatures}\label{appB}
Our QBA protocol can apply any kind of QDS to ensure unconditional security and better fault-tolerance performance.  

\subsection{BB84-KGP GC01-QDS}
\textcolor{black}{BB84-KGP GC01-QDS is a traditional single-bit QDS protocol proposed in 2016. In every round only one bit of message is signed. That is, possible message is $m$ = 0 or 1. In the distribution stage, bit correlations between Alice--Bob and Alice--Charlie are realized by BB84 key generation protocol (KGP).  In the messaging stage users exchange partial of their keys and compare the mismatch rate to verify the signature. Here we introduce this protocol used in our quantum consensus experiment.
}

\textcolor{black}{\textit{Distribution stage}---}

\textcolor{black}{(i) For $m$ = 0 or 1, Alice uses the BB84-KGP to generate four different keys of length $L$, $A_B^0,~A_B^1,~A_C^0,~A_C^1$, where the subscript $A$ and $B$ denotes she performed the KGP with Bob and Charlie, respectively, and the superscript denotes the future message to be signed, to be decided later by Alice. After BB84-KGP, Bob holds the length $L$ strings $K_B^0,~K_B^1$ and Charlie holds the length L strings $K_C^0,~K_C^1$.
	The procedure of BB84-KGP is analogous to BB84-QKD, but error correction and privacy amplification steps are removed. The shared keys are correlated with limited mismatch and secrecy leakage.}

\textcolor{black}{(ii) Bob and Charlie symmetrize their keys by choosing half of the bit values in their $K_B^m,~K_C^m$
	and sending them as well as the corresponding positions to each other using the Bob-Charlie secret classical channel. They will only keep the bits they did not forward and those received from the other participant. Their final symmetrized keys are denoted as $S_B^m$ and $S_C^m$. Bob (and Charlie) will keep a record of whether an element in $S_B^m$ ($S_C^m$) came directly from Alice or whether it was forwarded to him by Charlie (or Bob).
}

\textcolor{black}{\textit{Messaging stage}---}

\textcolor{black}{(i) To send a signed one-bit message $m$, Alice sends $(m,Sig_m)$ to the desired recipient (say Bob), where $sig_m=(A_B^m,A_C^m)$.}

\textcolor{black}{(ii) Bob checks whether $(m,Sig_m)$ matches his $S_B^m$ and records the number of mismatches he finds. He separately checks the part of his key received directly from Alice and the part of the key received from Charlie. If there are fewer than $s_a(L/2)$ mismatches in both halves of the key, where $s_a < 1/2$ is a small threshold determined by the parameters and the desired security level of the protocol, then Bob accepts the message.}

\textcolor{black}{(iii) To forward the message to Charlie, Bob forwards the pair $(m,Sig_m)$ that he received from Alice.}

\textcolor{black}{(iv) Charlie tests for mismatches in the same way, but in order to protect against repudiation by Alice he uses a different threshold. Charlie accepts the forwarded message if the number of mismatches in both halves of his key is below $s_v(L/2)$ where $s_v$ is another threshold, with $0 < s_a < s_v < 1/2$. }

\textcolor{black}{---\textit{Security of BB84-QDS}.
}

\textcolor{black}{The probability of a successful repudiation is }

\begin{equation}
	\textcolor{black}{\varepsilon_{\rm{rep}}=2e^{-(s_a-s_v)^2L},}
\end{equation}
\textcolor{black}{and that of a successful forgery is }
\begin{equation}
	\textcolor{black}{\varepsilon_{\rm{for}}=2e^{-\frac{1}{4}(p_e-s_a)^2L}, }
\end{equation}
\textcolor{black}{where $p_e$ represents the unknown information of one bit in the string and can be bounded by parameters of BB84-KGP.}

\subsection{One-time universal$_2$ hashing QDS}
We introduce the one-time universal$_2$ hashing (OTUH)-QDS we applied in our quantum consensus experiment, which utilizes secret sharing, one-time hashing and one-time pad to generate and verify signatures~\cite{yin2021experimental}. 

\textcolor{black}{\textit{Distributon stage}---}

Before executing the signature, Alice, Bob and Charlie all have two sets of keys, $X_{a,b,c}$ and $Y_{a,b,c}$, which satisfy the bit correlations $X_{a}=X_{b}\oplus X_{c}$ ($p$ bits) and $Y_{a}=Y_{b}\oplus Y_{c}$ ($2p$ bits). The perfect bit correlation of three parties can be realized by using quantum communication, such as quantum secret sharing and quantum key distribution.  \textcolor{black}{Note that OTUH-QDS requires that all three participants have the bit correlations $X_{a}=X_{b}\oplus X_{c}$ and $Y_{a}=Y_{b}\oplus Y_{c}$ before Alice signs the message, otherwise Bob and Charlie cannot successfully verify the signature.} In our experiment, we use four-intensity decoy-state BB84 QKD to implement this bit correlation. Alice shares the secret keys $X_b$ and $Y_b$ with Bob, and $X_c$ and $Y_c$ with Charlie via QKD. Then, Alice gets her own secret keys by XOR operation. Suppose that Alice signs a $q$-bit document (message), denoted as $m$, and sends it to `forwarder' Bob.

\textcolor{black}{\textit{Messaging stage}---}

(i) \emph{Signing}--Alice generates an irreducible polynomial~\cite{menezes2018handbook} $I(x)$ of degree $p$ at random using a local quantum random number, which can be characterized by an $p$-bit string $I_a$. Then she uses her key bit string $X_a$ and the irreducible polynomial $I(x)$ to generate a random linear feedback shift register-based (LFSR-based) Toeplitz matrix $H_{pq}$ of $p$ rows and $q$ columns. She acquires a $2p$-bit digest $Dig=(Dig_1||I_a)$. Here, $Dig_1$ is the digest of the $q$-bit document through a hash operation with $Dig_1$= $H_{pq} \cdot m$, and $I_a$ is an $p$-bit string for generating the irreducible polynomial in the LFSR-based Toeplitz matrix.
Then, Alice encrypts the digest with her key bit string $Y_{a}$ to obtain the $2p$-bit signature $Sig=Dig\oplus Y_{a}$. She sends the document and signature $\{Sig,~m\}$ to Bob.

(ii) \emph{Forwarding}-- Bob transmits $\{Sig,~m\}$ as well as his key bit strings $\{X_{b},~Y_{b}\}$ to Charlie to inform Charlie that he has received the signature. Then, Charlie forwards his key bit strings $\{X_{c},~Y_{c}\}$ to Bob. Bob obtains two new key bit strings $\{K_{X_{b}}=X_{b}\oplus X_{c},~K_{Y_{b}}=Y_{b}\oplus Y_{c}\}$ by the XOR operation.

(iii) \emph{Verification}-- Bob exploits $K_{Y_{b}}$ to obtain an expected digest and a string $I_b$ via XOR decryption. He utilizes $K_{X_{b}}$ and $I_b$ to establish an LFSR-based Toeplitz matrix and acquires an actual digest via a hash operation. Bob will accept the signature if the actual digest is equal to the expected digest. Then, he informs Charlie of the result. If Bob announces that he accepts the signature, Charlie creates two new key bit strings $\{K_{X_{c}}=X_{b}\oplus X_{c},~K_{Y_{c}}=Y_{b}\oplus Y_{c}\}$ using his original key and the key sent by Bob. He employs $K_{Y_{c}}$ to acquire an expected digest and a variable $I_c$ via XOR decryption. Charlie obtains an actual digest via a hash operation, where the hash function is an LFSR-based Toeplitz matrix generated by $K_{X_{c}}$ and $I_c$. Charlie accepts the signature if the two digests are identical.


\textcolor{black}{---\textit{Security of OTUH-QDS}. 
	This QDS protocol is naturally immune to repudiation and the probability of a successful forgery can be determined by 
	\begin{equation}
		\varepsilon_{\rm{for}}=\frac{|m|}{2^{p-1}},
	\end{equation}
	where $|m|$ is the length of the message. In this work, we choose $p=128$ and thus even for the $2^{64}$-bit document it is still safe enough.}

\subsection{OTUH-QDS without perfect keys with BB84-KGP}

\textcolor{black}{Recently, a variant of OTUH-QDS, called OTUH-QDS without perfect keys, was proposed~\cite{li2023one}. Different from OTUH-QDS that calls for sharing perfect quantum keys in the distribution stage, this variant users share keys through KGP which is consist with that in single-bit QDS. In the following we introduce OTUH-QDS without perfect keys with BB84-KGP that is used in our experiment demonstration.}

\textcolor{black}{
	\textit{Distribution stage}--- }

\textcolor{black}{
	(i) Alice-Bob and Alice-Charlie independently implement BB84-KGP to share correlated bit strings. This KGP process is the same as that in BB84-QDS. Thereafter, Alice-Bob and Alice-Charlie perform error correction algorithms on their shared bit strings. After this step, Alice hold two strings, denoted as $k_1^A$ and $k_2^A$. She obtains one string $k^A$ through XOR operation $k^A=k_1^A\oplus k_2^A$.  Bob and Charlie each holds one strings, denoted as $k^B$ and $k^C$, respectively.}

\textcolor{black}{
	(ii) Alice randomly disturbs the orders of $k^A$, and cuts the new string into $P$-bit subgroups. The size of $P$ is estimated by parameters of BB84-KGP so that the security is guaranteed. Alice will publicize the new order and $P$, and Bob and Charlie will perform the same operation on $k^B$ and $k^C$ accordingly.
}

\textcolor{black}{
	\textit{Messaging stage}--- }

\textcolor{black}{The messaging stage is analogous to that in OTUH-QDS. One subgroup in the distribution stage contributes $X_{a,b,c}$ with length $P$ and another two subgroups contribute $Y_{a,b,c}$ with length $2P$. The rules of Alice, Bob and Charlie are then consistent with that in OTUH-QDS.}

\textcolor{black}{---\textit{Security of OTUH-QDS without perfect keys}. }

\textcolor{black}{This protocol is also naturally immune to repudiation attacks. The probability of a successful forgery is limited by }
\begin{equation}
	\textcolor{black}{\varepsilon_{ \rm{for}}=|m|\cdot 2^{1-\mathcal{H}_P},}
\end{equation}
\textcolor{black}{where $\mathcal{H}_P$ is the unknown information of a $P$-bit subgroup generated in distribution stage, and can be estimated by parameters of BB84-KGP.}

\subsection{Colluding attack} \label{sec:colluding}
Colluding attacks are the most serious problem in decentralized quantum digital signatures involving multiple participants~\cite{Weng2021secure}. A colluding attack means that there are more than two malicious nodes colluding together to disturb the normal functioning of a system.
In our quantum Byzantine agreement, due to complete decentralization, colluding attacks appear as the number of malicious nodes increases. In a three-party QDS, if the sender and forwarder are dishonest, they can collude together to make another node believe the forged message and the corresponding signature. In the broadcasting phase of our QBA protocol, this will lead to inconsistency of the delivered messages of the two adjacent multicast rounds without prejudice to the rule of coherence, as we can see in Lemma~\ref{l1} and Lemma~\ref{l2}. It allows dishonest players to deliver inconsistent messages in the system only under colluding attacks.

\section{Security analysis}\label{appC}
In our QBA protocol, the performance of honest (dishonest) players follows the same rule. Therefore, the players can be divided into two groups, the honest and dishonest. Also, the elements of a gathering list can be divided in the same way. Therefore, we can simplify the protocol with a perfect binary tree model where one tree node represents the set of multicast rounds with honest or dishonest primaries. The left (right) child tree node represents the multicast rounds with honest (dishonest) primaries of the next depth. In what follows, when we say a tree node is honest(dishonest), it means that the primaries in this tree node is honest (dishonest). 
And we can obtain the important Lemma~\ref{l1} and Lemma~\ref{l2}.

\begin{lemma}\label{l1}
	Suppose that B is a right child tree node of a parent node A who is honest, and C is the left child tree node of B. The messages delivered in C are consistent with those of A, which protects the consistency of the delivered messages. 
\end{lemma}

\begin{figure}
	\centering
	\includegraphics[width=8.5cm]{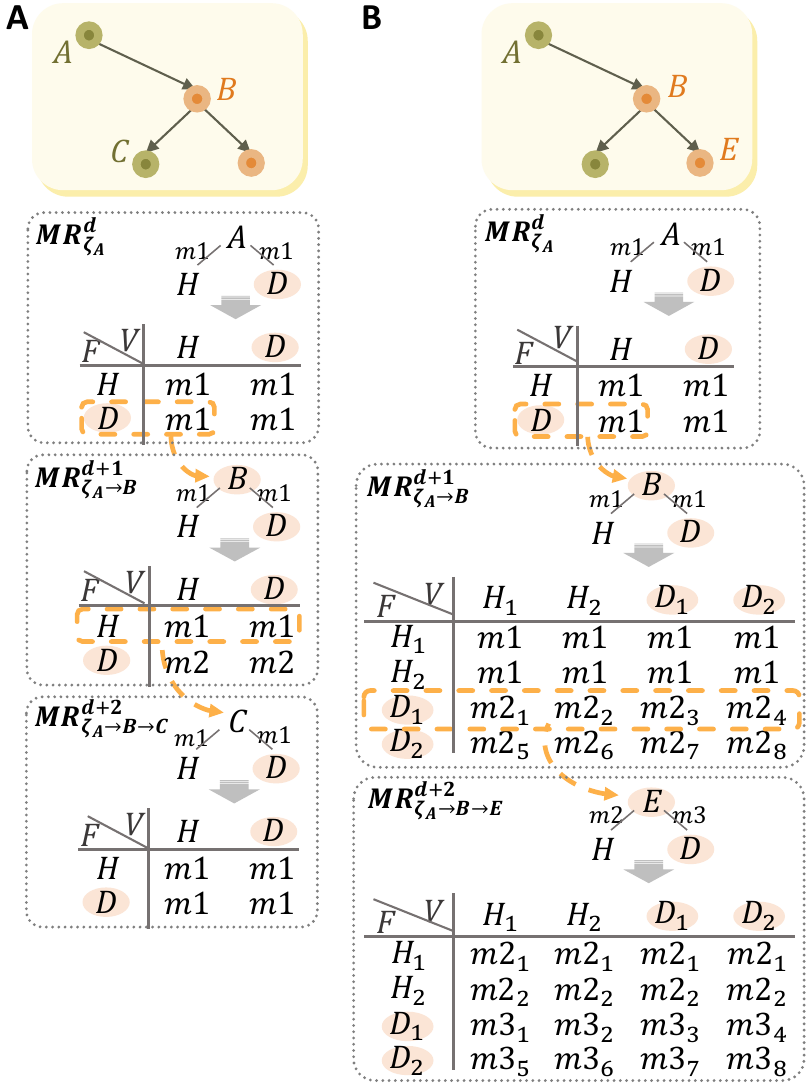}
	\caption{\textbf{Illustration of Lemma 1 and Lemma 2}. We use `F' to represent `forwarder', `V' to represent `verifier', `H' to represent `honest backups' and `D' to represent `dishonest backups' in the tables. The honest (dishonest) tree nodes are denoted by green (orange) nodes. The detailed process of the broadcasting phase is shown in tables where all recorded messages are summarized according to the honest and the dishonest case. (A) Illustration for Lemma 1. (B) Illustration for Lemma 2.}
	\label{sp1}
\end{figure}

\begin{proof}\label{p1}
	\rm As shown in Fig.~\ref{sp1}(A), each primary of the honest node A multicasts $m1$ to the backups. In this case, the dishonest forwarders cannot forward any messages except $m1$ due to the unforgeability of the QDS, and each verifier receives $m1$.
	Then, the coherence check guarantees that each primary of dishonest B must deliver $m1$ to the honest forwarders in $\mathcal{MR}^{d+1}_{\zeta_{A}\to B}$. Therefore, the messages multicast by the honest primaries of C in $\mathcal{MR}^{d+2}_{\zeta_{A}\to B\to C}$ are $m1$, where $\zeta_A$ marks the route before and containing A.
\end{proof}

\begin{lemma}\label{l2}
	Suppose that B is a right child tree node of a parent node A who is honest, and E is the right child tree node of B. The message multicast in E can be inconsistent with those of A, which disrupts the consistency of the delivered messages.
\end{lemma}

\begin{proof}\label{p2}
	\rm  As shown in Fig.~\ref{sp1}(B), each primary of the honest node A multicasts $m1$ to backups. Then, each primary of the dishonest node B can only deliver $m1$ to the honest backups. However, the primaries of the dishonest B can execute the colluding attack together with the dishonest forwarders, and they can deliver any conflicting messages to the verifiers. After forwarding the different messages, these dishonest forwarders in $\mathcal{MR}^{d+1}_{\zeta_{A}\to B}$ ($\zeta_A$ marks the route before and containing A), who are also the dishonest primaries of E, can multicast these conflicting messages without compromising consistency. The consistency of the delivered messages is completely disrupted.
\end{proof}

By Lemma 1 and Lemma 2, we find that on the route that avoids consecutively choosing the right child, the consistency of the delivered messages can be protected in the broadcasting phase. With this idea, we denote a special route as a safe path in the binary tree as follows.

\begin{figure*}
	\centering
	\includegraphics[width=17cm]{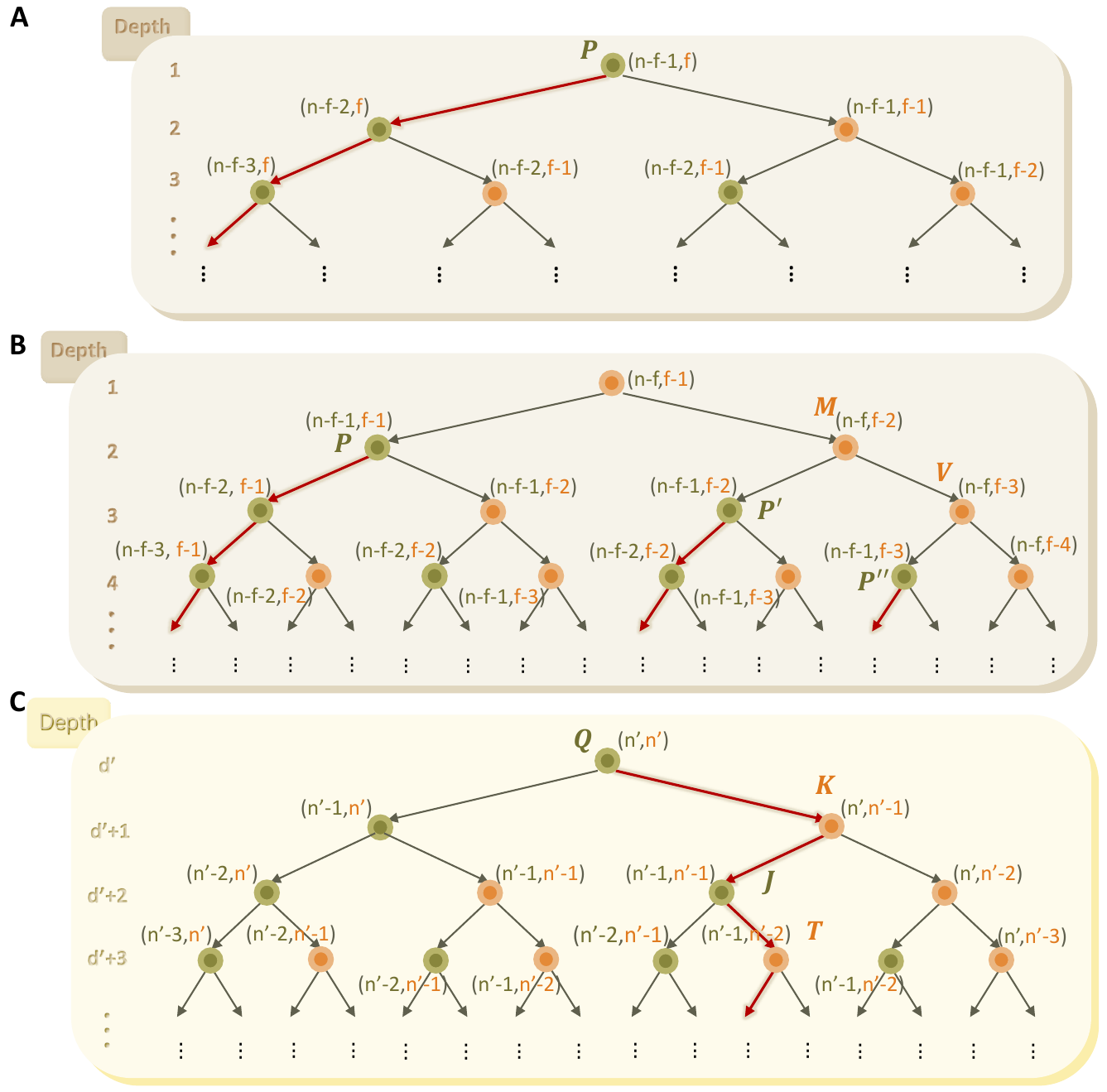}
	\caption{\textbf{The perfect binary tree model of our protocol and the safe path.} The green (orange) nodes in the tree represent the multicast rounds with honest (dishonest) primaries. Beside each node, the left number indicates the number of honest backups of this tree node, and the right number indicates the number of dishonest backups. The safe paths are represented by the red arrows. (A) and (B) illustrate the safe path before reaching the intermediate tree node. (A) The case in which the initial primary is honest. By the definition of a safe path, the honest initial primary is a safe node, denoted by P, and there is only one safe path in the whole process. (B) The case where the initial primary is dishonest and there is more than one safe path. The safe tree nodes are denoted by P, P$^\prime$, P$^{\prime\prime}$ and so on. (C) Illustration of the part of the safe path from the intermediate tree node Q to the ending tree node. The other tree nodes in the safe path are denoted by K, J, T and so on.}
	\label{sp2}
\end{figure*}

\begin{definition}\label{d1}
	\rm The safe tree node, denoted by P, is defined as the first honest tree node in the message delivery route from the top to the bottom layer, as shown in Fig.~\ref{sp2}(A). Note that in the safe tree node, at least half of the backups are honest. Then, we continuously choose the left child node layer-by-layer until we reach an intermediate tree node. The intermediate node is defined as the tree node that has an equivalent number of honest and dishonest backups of depth $d^\prime$, denoted by Q. As shown in Fig.~\ref{sp2}(C), from the intermediate tree node Q (depth $d^\prime$), we choose the right child K of Q ($d^\prime+1$), the left child J of K ($d^\prime+2$), the right child T of J ($d^\prime+3$), the right child of T ($d^\prime+4$), and so on. That is, the right and left child tree nodes are chosen in turn layer-by-layer until the ending tree node O of the penultimate depth is reached. This path from the safe tree node P, passing through the intermediate tree node Q, to reach the ending tree node O is defined as the safe path.
\end{definition}

\begin{lemma}
	The safe path ensures that the honest players in the safe tree node can reach consensus on their outputs.
\end{lemma}

\begin{figure}
	\centering
	\includegraphics[width=8.5cm]{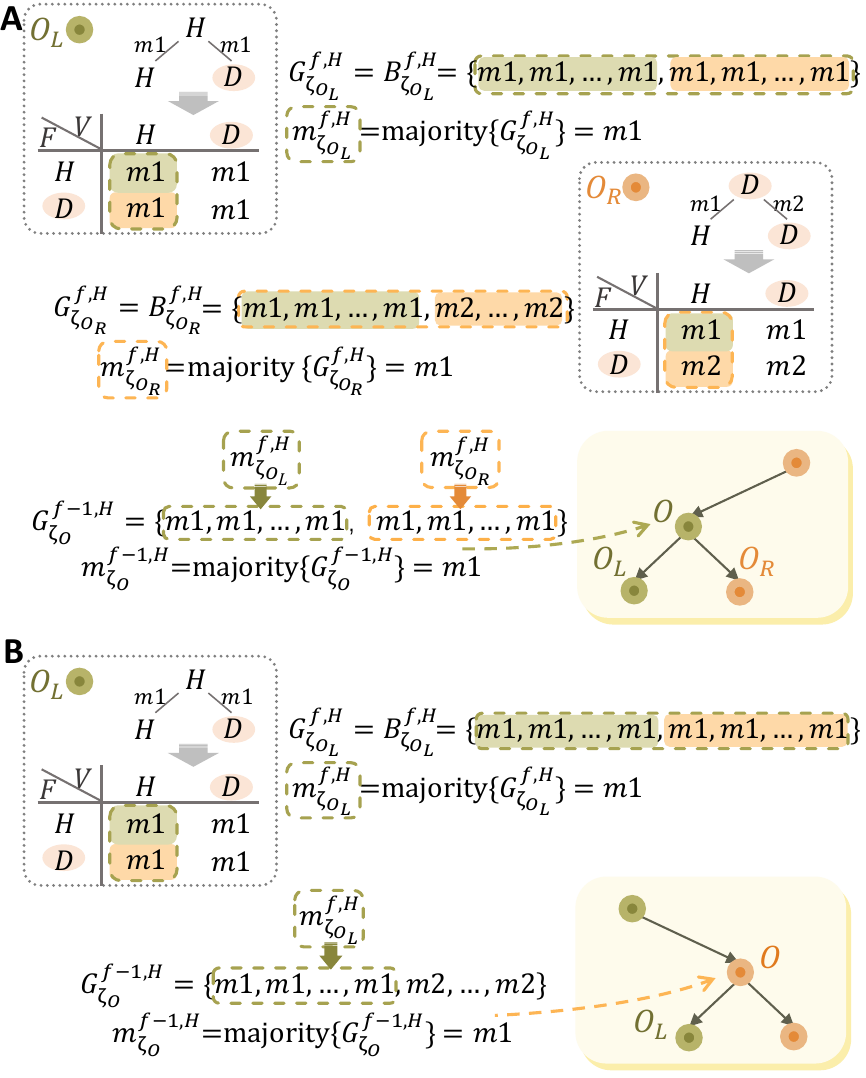}
	\caption{\textbf{The ending tree node of a safe path at the penultimate depth and its child tree nodes.} The yellow figure of the binary tree presents the ending tree node of a safe path, denoted by O, and its left (right) child tree node is denoted by O$_L$ (O$_R$). The messages contributed by the honest (dishonest) backups are marked by green (orange). The messages recorded in the broadcasting list are represented in the table. The gathering list and the corresponding outputs are also presented beside the tables.  (A) The ending tree node is honest. (B) The ending tree node is dishonest.}
	\label{sp3}
\end{figure}

\begin{proof}\label{p3}
	\rm In the broadcasting phase, by Lemma 1, the consistency of the message from the safe node's primaries can be protected. Considering one of the rounds of the safe tree node, the honest primary multicasts the message $m1$. For simplicity, the following discussion only analyses the route starting with this round. The analysis for the other rounds of the safe node is similar. On the safe path, the honest players in the subsequent rounds will receive and then multicast the message $m1$. 
	
	In the gathering phase, the message deducing process analyzed below demonstrates the consistency of the final outputs. From the safe tree node P to the intermediate tree node Q, more than half of the elements from the left (honest) child node appear in the gathering lists of each tree node. From the intermediate tree node Q to the ending tree node, if a tree node is honest, then in each gathering list of this node, the number of elements from the left child node is the same as that of the right child; if a tree node is dishonest, then in each gathering list of this node, the honest child node contributes one more element than the dishonest child node. Note that the output of each tree node in the safe path is determined by the tree nodes that are also on the safe path. Thus, all other branches in the binary tree can be ignored. 
	
	Our aim is to prove that in each tree node from the ending tree node to the safe tree node, more than half of the elements of each gathering list are always consistent, and thus the outputs of each node are always consistent during the recursion gathering process. We consider the two situations in Fig.~\ref{sp3}: (A) the ending tree node O is honest, and (B) the ending tree node O is dishonest. We denote the left (right) child tree node of O as O$_L$ (O$_R$).
	
	\textbf{(a). The ending tree node O is honest.} We first analyze the outputs from the initial gathering lists of O$_L$ and O$_R$. The message of O$_L$ is $m1$, which is also the message that is multicast in the safe node. In the bottom layer, each backup has the same gathering list where all the elements are $m1$, which is obtained directly from the bottom broadcasting list. Thus, the honest backups of O$_L$ have the same output. The primaries of O$_R$ multicast $m1$ to the honest backups. Since the honest backups contribute one element more than the dishonest backups in each gathering list, message $m1$ is the majority. Thus, each honest backup of O$_R$ has the same output $m1$. That is, the outputs from O$_L$ and O$_R$ are all $m1$. Then, the honest backups of the ending tree node O have the same output deduced from their consistent gathering lists. 
	
	Considering the consistent outputs of the ending tree node, consistency can always be held on the safe path. Suppose that node U is the dishonest parent tree node of O at $d = f-2$. The output of each gathering list of U is determined by elements from O. Thus, the outputs are also $m1$. On the safe path, the parent tree node of U has at least half of the elements, which are $m1$, in each gathering list. Moreover, for a certain backup's gathering list, there is also one element from the backup himself or herself, which is the message he or she received directly from the corresponding primary. This element is also $m1$ by Lemma 1. Therefore, more than half of the elements in each gathering list are $m1$, and the output is $m1$. Following the above process until the intermediate tree node Q is reached, we can see that all the outputs of the tree nodes on the safe path are $m1$. Finally, the honest backups of node Q have the same output $m1$.
	
	In the tree nodes from P to Q, for each gathering list, more than half of the elements are $m1$. Finally, in the safe tree node P, all honest players in each round will reach consensus.
	
	\textbf{(b). The ending tree node O is dishonest.} The output of O is determined by the elements from O$_L$. We find that these elements are all $m1$ since O$_L$ is honest. Thus, the backups in the ending tree node O have the same output $m1$. Similar to the analysis in (a), each backup of the intermediate node Q outputs $m1$. Therefore, all the honest players in the safe tree node P have consistent outputs $m1$.
\end{proof}

\bigskip
\noindent
\textbf{Theorem 1} For an $n$-player system with $f$ malicious players, our QBA protocol can reach consensus with a fault tolerance of $N\ge2f+1$.

\begin{proof}\label{p4}
	\rm We start with $N=2f+1$. We analyze it according to whether the initial primary is honest or dishonest.
	
	\textbf{(a). The initial primary is honest.} By Definition 1, the root tree node is not only the safe tree node but also the immediate tree node. By Lemma 3, all honest players in the initial round can reach consensus, which satisfies the Byzantine conditions IC$_1$ and IC$_2$.
	
	\textbf{(b). The initial primary is dishonest.} As shown in Fig.~\ref{sp2}(B), the dishonest initial primary S can arbitrarily deliver different messages to different forwarders at depth 1. At depth 2, the left (honest) child tree node is a safe tree node that starts a safe path. The primaries of the right (dishonest) child tree node M can execute colluding attacks and deliver conflicting messages as described in Lemma 2. The left child tree node P$^\prime$ of M is another safe tree node that starts another safe path. Similarly, the tree node P$^{\prime\prime}$ is also a safe tree node that starts another safe path, and so on.
	
	One of the honest player's outputs in the initial round is $m^1_S={\rm majority}(G^1_S)$, where $G^1_S=\left\{m_{S\to R_1},m_{S\to R_2} \cdots,m_{S \to R_{n-1}}\right\}$. By Proof~\ref{p3}, although the outputs of the tree node P may be different, the honest players reach consensus on each of these outputs since P is a safe tree node. Thus, in the gathering list ${G^1_S}$, the $f+1$ elements from the safe tree node P are consistent among the honest backups. Next, we discuss the $f-1$ elements of the list ${G^1_S}$ from node M, as shown in Fig.~\ref{sp2}(B). Similar to the above process, we can find that all honest backups reach consensus on the $f$ elements from the safe tree node P$^\prime$, so we must consider the $f-2$ elements of the list ${G^2_{S \to B}}$ from node V, and so on. After we continuously choose the right child tree node at the next depth, the dishonest backups of the tree nodes will continuously be reduced by one while the number of honest backups will not change. When we reach depth $f$, in the dishonest leaf node, only the primary is dishonest and all the backups are honest. 
	Thus, there are no colluding attacks. Although the outputs of different rounds may be different, the backups of the same round can obtain consistent outputs. 
	In the dishonest parent of this leaf node, each gathering list has one element from the this dishonest leaf node (right child node) and $f-1$ elements from the honest leaf node (left child node). 
	The left child node is also a safe node and these $f-1$ elements are also consistent. Thus each honest backup of this dishonest parent tree node also obtains a consistent output. Following the above recursion process, we find that the honest backups in each round of the above path always have the same gathering lists. 
	The consistent outputs from each safe node indirectly or directly lead to the eventual consistency of the elements that make up each gathering list of $d = 1$.
	In the initial round, the elements in a list may be different from each other, but the gathering lists of the honest players are the same, regardless of the messages delivered by the dishonest primary. 
	Finally, all the honest players reach consensus and output consistent messages, which satisfies the condition IC$_2$.
	
	In summary, we prove that our protocol can satisfy the two Byzantine conditions, IC$_1$ and IC$_2$, to reach Byzantine agreement when $N \ge 2f+1$.
\end{proof}

If $N\le 2f$, then the safe paths appear too late in the binary tree model. The consistency of the delivered messages cannot be guaranteed in the tree nodes before the safe path. For example, when $N=2f$ and the initial primary is honest. The root tree node is honest, but the number of honest backups is $f-1$ and the number of dishonest backups is $f$. Therefore, the root tree node is no longer a safe tree node. In fact, the minimum depth at which we can find a safe tree node is $d=4$ in the binary tree. There are two safe nodes, denoted as $P_1$ and $P_2$, that begin their safe paths at $d=4$. $P_1$ and $P_2$ can be found by the following steps. $P_1$: After choosing the right child tree node twice, the left tree node at depth 4 is $P_1$. $P_2$: First choosing the left child of the root tree node first, and then choosing the right tree node, finally the left tree node at depth 4 is $P_2$. By Lemma 2, the message multicast in $P_1$ can conflict with the message delivered by the initial primary. Suppose the messages multicast by the initial primary are $m1$ and the conflicting messages delivered by dishonest players are $m2$. Then the backups of tree node $P_1$ will consistently output $m2$. The backups of the tree node $P_2$ will still consistently output $m1$. After several rounds of counting, the numbers of messages $m1$ and $m2$ in his or her own gathering list for the initial round are $f-1$ and $f$, respectively. Then, all the honest backups in the initial round will output $m2$, while the honest initial primary outputs $m1$. Therefore, they cannot reach consensus. When $N < 2f$, the situation will undoubtedly worsen.

\section{Experiment calculation details}\label{appD}
\begin{figure*}
	\centering
	\includegraphics[width=16cm]{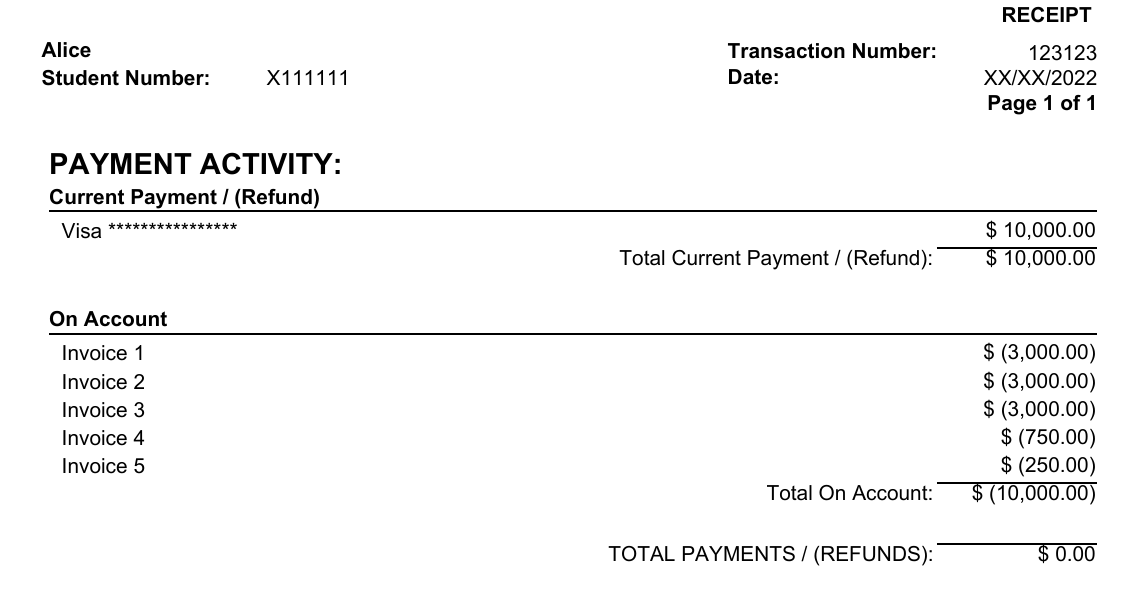}
	\caption{\textbf{The digital ledger for transmission in the experiment.} We convert the digital ledger into a binary string of bits. The binary string of bits are the actual message we transmit in the experiment. We denote the correct string as $m1$, and denote the conflicting messages delivered by dishonest players as $m2$, $m3$ and so on.}
	\label{receipt}
\end{figure*}

We experimentally implement the three-party consensus \textcolor{black}{utilizing GC01-QDS~\cite{amiri2016secure},  OTUH-QDS~\cite{yin2021experimental}, and OTUH-QDS without perfect keys~\cite{li2023one}, respectively, and implement the five-party consensus with OTUH-QDS.  Here, we utilize four-intensity decoy-state BB84 key generation process for the three QDS protocols~\cite{yin2020experimental}. } The decentralized digital ledger is shown in Fig.~\ref{receipt}, which is reached consensus on by the users in the experiment. The digital ledger is converted into a binary string of bits. We denote this correct message as $m1$, and the incorrect messages as $m2$, $m3$, and so on.

We will introduce the calculation details based on our experiment. 

$\overline{x}$ ($\underline{x}$) denotes the upper (lower) bound of the observed value $x$.
Using the decoy-state method for finite sample sizes, the expected number of vacuum events $\underline{s}_{0}^{z z^{\ast}}$ and single-photon 
events $\underline{s}_{1}^{z z^{\ast}}$ can be expressed as
\begin{equation}
	\underline{s}_{0}^{z z^{\ast}} \ge (e^{-\mu}p_{\mu}+e^{-\nu}p_{\nu})\frac{\underline{n}_{0}^{z^{\ast}}}{p_0},
\end{equation}
and
\begin{equation}
	\begin{aligned}
		\underline{s}_{1}^{z z^{\ast}} \ge &\frac{\mu^{2} e^{-\mu} p_{\mu}+\mu \nu e^{-\nu} p_{\nu}}{\mu \nu-\nu^{2}} \times \\
		&\left(e^{\nu} \frac{\underline{n}_{\nu}^{z^{\ast}}}{p_{\nu}}-\frac{\nu^{2}}{\mu^{2}} e^{\mu} \frac{\overline{n}_{\mu}^{z^{\ast}}}{p_{\mu}}-\frac{\mu^{2}-\nu^{2}}{\mu^{2}} \frac{\overline{n}_{0}^{z^{\ast}}}{p_{0}}\right),
	\end{aligned}
\end{equation}
respectively. Here $n_k^{z(x)}$ is the count of $k$ ($k\in \left\{\mu,\nu,\omega\right\}$) intensity pulse measured in the Z(X) basis, and $x^{\ast}$ is the expected value of observed value $x$. We use the variant of the Chernoff bound to obtain the lower and upper bounds, $\overline{x}^{\ast}=x+\beta+\sqrt{2\beta x+\beta^2}$ and $\underline{x}^{\ast}=x-\frac{\beta}{2}-\sqrt{2\beta x+\frac{\beta^2}{4}}$, where $\beta=\ln \frac{22}{\varepsilon}$.

The expected value of the number of single-photon events $\underline{s}_1^{xx^{\ast}}$ in $\chi_\omega$ can be given by
\begin{equation}
	\underline{s}_{1}^{x x^{*}} \ge \frac{\mu \omega e^{-\omega} p_{\omega}}{\mu \nu-\nu^{2}}\left(e^{\nu} \frac{\underline{n}_{\nu}^{x^{*}}}{p_{\nu}}-\frac{\nu^{2}}{\mu^{2}} e^{\mu} \frac{\overline{n}_{\mu}^{x^{*}}}{p_{\mu}}-\frac{\mu^{2}-\nu^{2}}{\mu^{2}} \frac{\overline{n}_{0}^{x^{*}}}{p_{0}}\right).
\end{equation}

Additionally, the expected number of bit errors $\underline{t}_1^{xx^{\ast}}$ associated with the single-photon event in $\chi_\omega$ is $\overline{t}_1^{xx} \le m_{\omega}^{x}-\underline{t}_0^{xx}$, where $\underline{t}_0^{xx}=\frac{e^{-\omega}p_{\omega}}{2p_0}\underline{n}_0^{x^{\ast}}$. For a given expected value $x^{\ast}$, the upper and lower bounds of the observed value are given by $\overline{x}=x^{\ast}+\frac{\beta}{2}+\sqrt{2\beta x^{\ast}+ \frac{\beta^2}{4}}$ and $\underline{x}=x^{\ast}-\sqrt{2\beta x^{\ast}}$, respectively. Using random sampling without replacement, the phase error rate in the Z basis is
\begin{equation}
	\overline{\phi}_{1}^{z z}=\frac{\overline{t}_{1}^{x x}}{\underline{s}_{1}^{x x}}+\gamma^{U}\left(\underline{s}_{1}^{z z}, \underline{s}_{1}^{x x}, \frac{\overline{t}_{1}^{x x}}{\underline{s}_{1}^{x x}}, \frac{\varepsilon}{22}\right),
\end{equation}
where $\gamma^{U}(n, k, \lambda, \epsilon)=\frac{\frac{(1-2 \lambda) A G}{n+k}+\sqrt{\frac{A^{2} G^{2}}{(n+k)^{2}}+4 \lambda(1-\lambda) G}}{2+2 \frac{A^{2} G}{(n+k)^{2}}}$, $A=\max\left\{n,k\right\}$ and $G=\frac{n+k}{n k} \ln \frac{n+k}{2 \pi n k \lambda(1-\lambda) \epsilon^{2}}$.

\subsubsection{Single-bit  GC01-QDS}
\textcolor{black}{In BB84-KGP GC01-QDS, the unknown information  to the attacker is given by}
\begin{equation}
	\textcolor{black}{\mathcal{H}= \underline{s}_{0}^{z z}  + \underline{s}_{1}^{zz}(1-h(\overline{\phi}_{1}^{zz}))},
\end{equation}
\textcolor{black}{where $h(x):=-x\log _{2} x-(1-x)\log_{2} (1-x)$.
	According to the $\mathcal{H}$, we can obtain the
	signature rate }
\begin{equation}
	\textcolor{black}{ SR =\frac{k}{|m|\cdot 2N},}
\end{equation}
\textcolor{black}{where $k=96$ MHz is the effective repetition rate, $|m|$ is the length of the message, $2N$ is the minimum number of pulses required to securely sign a one-bit message according to the set security parameter.}

\subsubsection{OTUH-QDS}

The length of the final key, which is $\varepsilon_{cor}$-correct and $\varepsilon_{sec}$-secret, can be expressed by~\cite{yin2020experimental}
\begin{equation}
	\ell=\underline{s}_{0}^{z z}+\underline{s}_{1}^{z z}\left[1-h\left(\overline{\phi}_{1}^{z z}\right)\right]-\lambda_{\mathrm{EC}}-\log _{2} \frac{2}{\varepsilon_{\mathrm{cor}}}-6 \log _{2} \frac{22}{\varepsilon_{\mathrm{sec}}},
\end{equation}
\textcolor{black}{and the signature rate of BB84-KGP OTUH-QDS can be expressed as}
\begin{equation}
	\textcolor{black}{SR=\frac{l}{3p\cdot t},}
\end{equation}
\textcolor{black}{where $p=128$ and $t$ is the time of sending pulses using 96 MHZ repetition rate.}

\subsubsection{OTUH-QDS without perfect keys}
\textcolor{black}{In OTUH-QDS without perfect keys based on BB84-KGP, Alice and Bob (Alice and Charlie) form the $n_Z$-length raw key bit from the random bits under the $Z$ basis. We can estimate parameters in a selected $P$-bit group, i.e.,
	the lower bound of number of vacuum events and single-photon events under the $Z$ basis $\underline{s}_{0,P}^{z z}$ and $\underline{s}_{1,P}^{z z}$, and the upper bound of the phase error rate of the single-photon events in the $Z$ basis $\overline{\phi}_{1,P}^{zz}$.}
\begin{equation}
	\textcolor{black}{\underline{s}_{0,P}^{z z}\geq P\left[\underline{s}_{0}^{zz}/n_Z-	\gamma^U (P,n_Z-P,\underline{s}_{0}^{zz}/n_Z,\varepsilon)\right],}
\end{equation}

\begin{equation}
	\textcolor{black}{\underline{s}_{1,P}^{z z}\geq P\left[\underline{s}_{1}^{zz}/n_Z-	\gamma^U (P,n_Z-P,\underline{s}_{1}^{zz}/n_Z,\varepsilon)\right],}
\end{equation}

\begin{equation}
	\textcolor{black}{\overline{\phi}_{1,P}^{zz}\leq \overline{\phi}_{1}^{zz}+\gamma^U \left(\underline{s}_{1,P}^{zz},\underline{s}_{1}^{zz}-\underline{s}_{1,P}^{zz},\overline{\phi}_{1}^{zz},\varepsilon\right).}
\end{equation}
\textcolor{black}{Finally we can obtain the unknown information of the $P$-bit group~\cite{li2023one}}
\begin{equation}
	\textcolor{black}{\mathcal{H}_P= \underline{s}_{0,P}^{z z} + \underline{s}_{1,P}^{z z}\left[1-h(\overline{\phi}_{1,P}^{zz})\right]-\lambda_{EC}.}
\end{equation}
\textcolor{black}{According to the $\mathcal{H}_P$ and the set security parameter, we can obtain the signature rate}
\begin{equation}
	\textcolor{black}{SR=\frac{n_Z}{P},}
\end{equation}
\textcolor{black}{where $P$ is the minimum number with the condition}
\begin{equation}
	\textcolor{black}{\epsilon_{\rm{for}}=|m|\cdot 2^{1-\mathcal{H}_P}\le \epsilon}
\end{equation}
\textcolor{black}{satisfied, where $\epsilon$ is upper bound of the failure probability of the QDS protocol.}


\begin{thebibliography}{70}%
\makeatletter
\providecommand \@ifxundefined [1]{%
 \@ifx{#1\undefined}
}%
\providecommand \@ifnum [1]{%
 \ifnum #1\expandafter \@firstoftwo
 \else \expandafter \@secondoftwo
 \fi
}%
\providecommand \@ifx [1]{%
 \ifx #1\expandafter \@firstoftwo
 \else \expandafter \@secondoftwo
 \fi
}%
\providecommand \natexlab [1]{#1}%
\providecommand \enquote  [1]{``#1''}%
\providecommand \bibnamefont  [1]{#1}%
\providecommand \bibfnamefont [1]{#1}%
\providecommand \citenamefont [1]{#1}%
\providecommand \href@noop [0]{\@secondoftwo}%
\providecommand \href [0]{\begingroup \@sanitize@url \@href}%
\providecommand \@href[1]{\@@startlink{#1}\@@href}%
\providecommand \@@href[1]{\endgroup#1\@@endlink}%
\providecommand \@sanitize@url [0]{\catcode `\\12\catcode `\$12\catcode `\&12\catcode `\#12\catcode `\^12\catcode `\_12\catcode `\%12\relax}%
\providecommand \@@startlink[1]{}%
\providecommand \@@endlink[0]{}%
\providecommand \url  [0]{\begingroup\@sanitize@url \@url }%
\providecommand \@url [1]{\endgroup\@href {#1}{\urlprefix }}%
\providecommand \urlprefix  [0]{URL }%
\providecommand \Eprint [0]{\href }%
\providecommand \doibase [0]{https://doi.org/}%
\providecommand \selectlanguage [0]{\@gobble}%
\providecommand \bibinfo  [0]{\@secondoftwo}%
\providecommand \bibfield  [0]{\@secondoftwo}%
\providecommand \translation [1]{[#1]}%
\providecommand \BibitemOpen [0]{}%
\providecommand \bibitemStop [0]{}%
\providecommand \bibitemNoStop [0]{.\EOS\space}%
\providecommand \EOS [0]{\spacefactor3000\relax}%
\providecommand \BibitemShut  [1]{\csname bibitem#1\endcsname}%
\let\auto@bib@innerbib\@empty
\bibitem [{\citenamefont {Lamport}\ \emph {et~al.}(1982)\citenamefont {Lamport}, \citenamefont {Shostak},\ and\ \citenamefont {Pease}}]{lamport1982byzantine}%
  \BibitemOpen
  \bibfield  {author} {\bibinfo {author} {\bibfnamefont {L.}~\bibnamefont {Lamport}}, \bibinfo {author} {\bibfnamefont {R.}~\bibnamefont {Shostak}},\ and\ \bibinfo {author} {\bibfnamefont {M.}~\bibnamefont {Pease}},\ }\bibfield  {title} {\bibinfo {title} {The {B}yzantine generals problem},\ }\href@noop {} {\bibfield  {journal} {\bibinfo  {journal} {ACM Transactions on Programming Languages and Systems}\ }\textbf {\bibinfo {volume} {4}},\ \bibinfo {pages} {382} (\bibinfo {year} {1982})}\BibitemShut {NoStop}%
\bibitem [{\citenamefont {Extance}(2015)}]{Extance2015future}%
  \BibitemOpen
  \bibfield  {author} {\bibinfo {author} {\bibfnamefont {A.}~\bibnamefont {Extance}},\ }\bibfield  {title} {\bibinfo {title} {The future of cryptocurrencies: Bitcoin and beyond},\ }\href@noop {} {\bibfield  {journal} {\bibinfo  {journal} {Nature}\ }\textbf {\bibinfo {volume} {526}},\ \bibinfo {pages} {21} (\bibinfo {year} {2015})}\BibitemShut {NoStop}%
\bibitem [{\citenamefont {Castro}\ \emph {et~al.}(1999)\citenamefont {Castro}, \citenamefont {Liskov} \emph {et~al.}}]{castro1999practical}%
  \BibitemOpen
  \bibfield  {author} {\bibinfo {author} {\bibfnamefont {M.}~\bibnamefont {Castro}}, \bibinfo {author} {\bibfnamefont {B.}~\bibnamefont {Liskov}}, \emph {et~al.},\ }\bibfield  {title} {\bibinfo {title} {Practical {B}yzantine fault tolerance},\ }in\ \href@noop {} {\emph {\bibinfo {booktitle} {OSDI}}},\ Vol.~\bibinfo {volume} {99}\ (\bibinfo {year} {1999})\ pp.\ \bibinfo {pages} {173--186}\BibitemShut {NoStop}%
\bibitem [{\citenamefont {Castro}\ and\ \citenamefont {Liskov}(2002)}]{Castro2002practical}%
  \BibitemOpen
  \bibfield  {author} {\bibinfo {author} {\bibfnamefont {M.}~\bibnamefont {Castro}}\ and\ \bibinfo {author} {\bibfnamefont {B.}~\bibnamefont {Liskov}},\ }\bibfield  {title} {\bibinfo {title} {Practical {B}yzantine fault tolerance and proactive recovery},\ }\href@noop {} {\bibfield  {journal} {\bibinfo  {journal} {ACM Trans. Comput. Syst.}\ }\textbf {\bibinfo {volume} {20}},\ \bibinfo {pages} {398–461} (\bibinfo {year} {2002})}\BibitemShut {NoStop}%
\bibitem [{\citenamefont {Aublin}\ \emph {et~al.}(2013)\citenamefont {Aublin}, \citenamefont {Mokhtar},\ and\ \citenamefont {Quéma}}]{aublin2013RBFT}%
  \BibitemOpen
  \bibfield  {author} {\bibinfo {author} {\bibfnamefont {P.-L.}\ \bibnamefont {Aublin}}, \bibinfo {author} {\bibfnamefont {S.~B.}\ \bibnamefont {Mokhtar}},\ and\ \bibinfo {author} {\bibfnamefont {V.}~\bibnamefont {Quéma}},\ }\bibfield  {title} {\bibinfo {title} {{RBFT}: Redundant {B}yzantine fault tolerance},\ }in\ \href@noop {} {\emph {\bibinfo {booktitle} {2013 IEEE 33rd International Conference on Distributed Computing Systems}}}\ (\bibinfo {year} {2013})\ pp.\ \bibinfo {pages} {297--306}\BibitemShut {NoStop}%
\bibitem [{\citenamefont {Miller}\ \emph {et~al.}(2016)\citenamefont {Miller}, \citenamefont {Xia}, \citenamefont {Croman}, \citenamefont {Shi},\ and\ \citenamefont {Song}}]{miller2016HBBFT}%
  \BibitemOpen
  \bibfield  {author} {\bibinfo {author} {\bibfnamefont {A.}~\bibnamefont {Miller}}, \bibinfo {author} {\bibfnamefont {Y.}~\bibnamefont {Xia}}, \bibinfo {author} {\bibfnamefont {K.}~\bibnamefont {Croman}}, \bibinfo {author} {\bibfnamefont {E.}~\bibnamefont {Shi}},\ and\ \bibinfo {author} {\bibfnamefont {D.}~\bibnamefont {Song}},\ }\bibfield  {title} {\bibinfo {title} {The honey badger of {BFT} protocols},\ }in\ \href@noop {} {\emph {\bibinfo {booktitle} {Proceedings of the 2016 ACM SIGSAC Conference on Computer and Communications Security}}},\ \bibinfo {series and number} {CCS '16}\ (\bibinfo  {publisher} {Association for Computing Machinery},\ \bibinfo {year} {2016})\ p.\ \bibinfo {pages} {31–42}\BibitemShut {NoStop}%
\bibitem [{\citenamefont {Yin}\ \emph {et~al.}(2019)\citenamefont {Yin}, \citenamefont {Malkhi}, \citenamefont {Reiter}, \citenamefont {Gueta},\ and\ \citenamefont {Abraham}}]{yin2019hotstuff}%
  \BibitemOpen
  \bibfield  {author} {\bibinfo {author} {\bibfnamefont {M.}~\bibnamefont {Yin}}, \bibinfo {author} {\bibfnamefont {D.}~\bibnamefont {Malkhi}}, \bibinfo {author} {\bibfnamefont {M.~K.}\ \bibnamefont {Reiter}}, \bibinfo {author} {\bibfnamefont {G.~G.}\ \bibnamefont {Gueta}},\ and\ \bibinfo {author} {\bibfnamefont {I.}~\bibnamefont {Abraham}},\ }\bibfield  {title} {\bibinfo {title} {Hotstuff: {BFT} consensus with linearity and responsiveness},\ }in\ \href@noop {} {\emph {\bibinfo {booktitle} {Proceedings of the 2019 ACM Symposium on Principles of Distributed Computing}}},\ \bibinfo {series and number} {PODC '19}\ (\bibinfo  {publisher} {Association for Computing Machinery},\ \bibinfo {year} {2019})\ p.\ \bibinfo {pages} {347–356}\BibitemShut {NoStop}%
\bibitem [{\citenamefont {Guo}\ \emph {et~al.}(2020)\citenamefont {Guo}, \citenamefont {Lu}, \citenamefont {Tang}, \citenamefont {Xu},\ and\ \citenamefont {Zhang}}]{guo2020dumbo}%
  \BibitemOpen
  \bibfield  {author} {\bibinfo {author} {\bibfnamefont {B.}~\bibnamefont {Guo}}, \bibinfo {author} {\bibfnamefont {Z.}~\bibnamefont {Lu}}, \bibinfo {author} {\bibfnamefont {Q.}~\bibnamefont {Tang}}, \bibinfo {author} {\bibfnamefont {J.}~\bibnamefont {Xu}},\ and\ \bibinfo {author} {\bibfnamefont {Z.}~\bibnamefont {Zhang}},\ }\bibfield  {title} {\bibinfo {title} {Dumbo: Faster asynchronous {BFT} protocols},\ }in\ \href@noop {} {\emph {\bibinfo {booktitle} {Proceedings of the 2020 ACM SIGSAC Conference on Computer and Communications Security}}},\ \bibinfo {series and number} {CCS '20}\ (\bibinfo  {publisher} {Association for Computing Machinery},\ \bibinfo {year} {2020})\ p.\ \bibinfo {pages} {803–818}\BibitemShut {NoStop}%
\bibitem [{\citenamefont {Lu}\ \emph {et~al.}(2020)\citenamefont {Lu}, \citenamefont {Lu}, \citenamefont {Tang},\ and\ \citenamefont {Wang}}]{lu2020dumbo}%
  \BibitemOpen
  \bibfield  {author} {\bibinfo {author} {\bibfnamefont {Y.}~\bibnamefont {Lu}}, \bibinfo {author} {\bibfnamefont {Z.}~\bibnamefont {Lu}}, \bibinfo {author} {\bibfnamefont {Q.}~\bibnamefont {Tang}},\ and\ \bibinfo {author} {\bibfnamefont {G.}~\bibnamefont {Wang}},\ }\bibfield  {title} {\bibinfo {title} {Dumbo-{MVBA}: Optimal multi-valued validated asynchronous {B}yzantine agreement, revisited},\ }in\ \href@noop {} {\emph {\bibinfo {booktitle} {Proceedings of the 39th Symposium on Principles of Distributed Computing}}},\ \bibinfo {series and number} {PODC '20}\ (\bibinfo  {publisher} {Association for Computing Machinery},\ \bibinfo {year} {2020})\ p.\ \bibinfo {pages} {129–138}\BibitemShut {NoStop}%
\bibitem [{\citenamefont {Pease}\ \emph {et~al.}(1980)\citenamefont {Pease}, \citenamefont {Shostak},\ and\ \citenamefont {Lamport}}]{pease1980reaching}%
  \BibitemOpen
  \bibfield  {author} {\bibinfo {author} {\bibfnamefont {M.}~\bibnamefont {Pease}}, \bibinfo {author} {\bibfnamefont {R.}~\bibnamefont {Shostak}},\ and\ \bibinfo {author} {\bibfnamefont {L.}~\bibnamefont {Lamport}},\ }\bibfield  {title} {\bibinfo {title} {Reaching agreement in the presence of faults},\ }\href@noop {} {\bibfield  {journal} {\bibinfo  {journal} {Journal of the ACM}\ }\textbf {\bibinfo {volume} {27}},\ \bibinfo {pages} {228} (\bibinfo {year} {1980})}\BibitemShut {NoStop}%
\bibitem [{\citenamefont {Dolev}\ \emph {et~al.}(1986)\citenamefont {Dolev}, \citenamefont {Halpern},\ and\ \citenamefont {Strong}}]{dolev1986possibility}%
  \BibitemOpen
  \bibfield  {author} {\bibinfo {author} {\bibfnamefont {D.}~\bibnamefont {Dolev}}, \bibinfo {author} {\bibfnamefont {J.~Y.}\ \bibnamefont {Halpern}},\ and\ \bibinfo {author} {\bibfnamefont {H.~R.}\ \bibnamefont {Strong}},\ }\bibfield  {title} {\bibinfo {title} {On the possibility and impossibility of achieving clock synchronization},\ }\href@noop {} {\bibfield  {journal} {\bibinfo  {journal} {Journal of Computer and System Sciences}\ }\textbf {\bibinfo {volume} {32}},\ \bibinfo {pages} {230} (\bibinfo {year} {1986})}\BibitemShut {NoStop}%
\bibitem [{\citenamefont {Fischer}\ \emph {et~al.}(1986)\citenamefont {Fischer}, \citenamefont {Lynch},\ and\ \citenamefont {Merritt}}]{fischer1986easy}%
  \BibitemOpen
  \bibfield  {author} {\bibinfo {author} {\bibfnamefont {M.~J.}\ \bibnamefont {Fischer}}, \bibinfo {author} {\bibfnamefont {N.~A.}\ \bibnamefont {Lynch}},\ and\ \bibinfo {author} {\bibfnamefont {M.}~\bibnamefont {Merritt}},\ }\bibfield  {title} {\bibinfo {title} {Easy impossibility proofs for distributed consensus problems},\ }\href@noop {} {\bibfield  {journal} {\bibinfo  {journal} {Distributed Computing}\ }\textbf {\bibinfo {volume} {1}},\ \bibinfo {pages} {26} (\bibinfo {year} {1986})}\BibitemShut {NoStop}%
\bibitem [{\citenamefont {Fitzi}\ \emph {et~al.}(2001{\natexlab{a}})\citenamefont {Fitzi}, \citenamefont {Garay}, \citenamefont {Maurer},\ and\ \citenamefont {Ostrovsky}}]{fitzi2001advances}%
  \BibitemOpen
  \bibfield  {author} {\bibinfo {author} {\bibfnamefont {M.}~\bibnamefont {Fitzi}}, \bibinfo {author} {\bibfnamefont {J.}~\bibnamefont {Garay}}, \bibinfo {author} {\bibfnamefont {U.}~\bibnamefont {Maurer}},\ and\ \bibinfo {author} {\bibfnamefont {R.}~\bibnamefont {Ostrovsky}},\ }\bibfield  {title} {\bibinfo {title} {Advances in cryptology-crypto 2001: Proceedings of the 21-st annual international cryptology conference, {S}anta {B}arbara, {CA}, 2001},\ }\href@noop {} {\bibfield  {journal} {\bibinfo  {journal} {Lecture Notes in Computer Science (Springer, Berlin, New York, 2001)}\ } (\bibinfo {year} {2001}{\natexlab{a}})}\BibitemShut {NoStop}%
\bibitem [{\citenamefont {Kiktenko}\ \emph {et~al.}(2018)\citenamefont {Kiktenko}, \citenamefont {Pozhar}, \citenamefont {Anufriev}, \citenamefont {Trushechkin}, \citenamefont {Yunusov}, \citenamefont {Kurochkin}, \citenamefont {Lvovsky},\ and\ \citenamefont {Fedorov}}]{kiktenko2018quantum}%
  \BibitemOpen
  \bibfield  {author} {\bibinfo {author} {\bibfnamefont {E.~O.}\ \bibnamefont {Kiktenko}}, \bibinfo {author} {\bibfnamefont {N.~O.}\ \bibnamefont {Pozhar}}, \bibinfo {author} {\bibfnamefont {M.~N.}\ \bibnamefont {Anufriev}}, \bibinfo {author} {\bibfnamefont {A.~S.}\ \bibnamefont {Trushechkin}}, \bibinfo {author} {\bibfnamefont {R.~R.}\ \bibnamefont {Yunusov}}, \bibinfo {author} {\bibfnamefont {Y.~V.}\ \bibnamefont {Kurochkin}}, \bibinfo {author} {\bibfnamefont {A.}~\bibnamefont {Lvovsky}},\ and\ \bibinfo {author} {\bibfnamefont {A.~K.}\ \bibnamefont {Fedorov}},\ }\bibfield  {title} {\bibinfo {title} {Quantum-secured blockchain},\ }\href@noop {} {\bibfield  {journal} {\bibinfo  {journal} {Quantum Sci. Technol.}\ }\textbf {\bibinfo {volume} {3}},\ \bibinfo {pages} {035004} (\bibinfo {year} {2018})}\BibitemShut {NoStop}%
\bibitem [{\citenamefont {Menezes}\ \emph {et~al.}(2018)\citenamefont {Menezes}, \citenamefont {Van~Oorschot},\ and\ \citenamefont {Vanstone}}]{menezes2018handbook}%
  \BibitemOpen
  \bibfield  {author} {\bibinfo {author} {\bibfnamefont {A.~J.}\ \bibnamefont {Menezes}}, \bibinfo {author} {\bibfnamefont {P.~C.}\ \bibnamefont {Van~Oorschot}},\ and\ \bibinfo {author} {\bibfnamefont {S.~A.}\ \bibnamefont {Vanstone}},\ }\href@noop {} {\emph {\bibinfo {title} {Handbook of applied cryptography}}}\ (\bibinfo  {publisher} {CRC press},\ \bibinfo {year} {2018})\BibitemShut {NoStop}%
\bibitem [{\citenamefont {Shor}(1994)}]{shor1994algorithms}%
  \BibitemOpen
  \bibfield  {author} {\bibinfo {author} {\bibfnamefont {P.}~\bibnamefont {Shor}},\ }\bibfield  {title} {\bibinfo {title} {Algorithms for quantum computation: discrete logarithms and factoring},\ }in\ \href@noop {} {\emph {\bibinfo {booktitle} {Proceedings 35th Annual Symposium on Foundations of Computer Science}}}\ (\bibinfo {year} {1994})\ pp.\ \bibinfo {pages} {124--134}\BibitemShut {NoStop}%
\bibitem [{\citenamefont {Grover}(1997)}]{grover1997quantum}%
  \BibitemOpen
  \bibfield  {author} {\bibinfo {author} {\bibfnamefont {L.~K.}\ \bibnamefont {Grover}},\ }\bibfield  {title} {\bibinfo {title} {Quantum mechanics helps in searching for a needle in a haystack},\ }\href@noop {} {\bibfield  {journal} {\bibinfo  {journal} {Phys. Rev. Lett.}\ }\textbf {\bibinfo {volume} {79}},\ \bibinfo {pages} {325} (\bibinfo {year} {1997})}\BibitemShut {NoStop}%
\bibitem [{\citenamefont {Arute}\ \emph {et~al.}(2019)\citenamefont {Arute}, \citenamefont {Arya}, \citenamefont {Babbush}, \citenamefont {Bacon}, \citenamefont {Bardin}, \citenamefont {Barends}, \citenamefont {Biswas}, \citenamefont {Boixo}, \citenamefont {Brandao}, \citenamefont {Buell} \emph {et~al.}}]{arute2019quantum}%
  \BibitemOpen
  \bibfield  {author} {\bibinfo {author} {\bibfnamefont {F.}~\bibnamefont {Arute}}, \bibinfo {author} {\bibfnamefont {K.}~\bibnamefont {Arya}}, \bibinfo {author} {\bibfnamefont {R.}~\bibnamefont {Babbush}}, \bibinfo {author} {\bibfnamefont {D.}~\bibnamefont {Bacon}}, \bibinfo {author} {\bibfnamefont {J.~C.}\ \bibnamefont {Bardin}}, \bibinfo {author} {\bibfnamefont {R.}~\bibnamefont {Barends}}, \bibinfo {author} {\bibfnamefont {R.}~\bibnamefont {Biswas}}, \bibinfo {author} {\bibfnamefont {S.}~\bibnamefont {Boixo}}, \bibinfo {author} {\bibfnamefont {F.~G.}\ \bibnamefont {Brandao}}, \bibinfo {author} {\bibfnamefont {D.~A.}\ \bibnamefont {Buell}}, \emph {et~al.},\ }\bibfield  {title} {\bibinfo {title} {Quantum supremacy using a programmable superconducting processor},\ }\href@noop {} {\bibfield  {journal} {\bibinfo  {journal} {Nature}\ }\textbf {\bibinfo {volume} {574}},\ \bibinfo {pages} {505} (\bibinfo {year} {2019})}\BibitemShut {NoStop}%
\bibitem [{\citenamefont {Fedorov}\ \emph {et~al.}(2018)\citenamefont {Fedorov}, \citenamefont {Kiktenko},\ and\ \citenamefont {Lvovsky}}]{fedorov2018quantum}%
  \BibitemOpen
  \bibfield  {author} {\bibinfo {author} {\bibfnamefont {A.~K.}\ \bibnamefont {Fedorov}}, \bibinfo {author} {\bibfnamefont {E.~O.}\ \bibnamefont {Kiktenko}},\ and\ \bibinfo {author} {\bibfnamefont {A.~I.}\ \bibnamefont {Lvovsky}},\ }\bibfield  {title} {\bibinfo {title} {Quantum computers put blockchain security at risk},\ }\href@noop {} {\bibfield  {journal} {\bibinfo  {journal} {Nature}\ }\textbf {\bibinfo {volume} {563}},\ \bibinfo {pages} {465} (\bibinfo {year} {2018})}\BibitemShut {NoStop}%
\bibitem [{\citenamefont {Wei}\ \emph {et~al.}(2020)\citenamefont {Wei}, \citenamefont {Li},\ and\ \citenamefont {Long}}]{wei2020full}%
  \BibitemOpen
  \bibfield  {author} {\bibinfo {author} {\bibfnamefont {S.}~\bibnamefont {Wei}}, \bibinfo {author} {\bibfnamefont {H.}~\bibnamefont {Li}},\ and\ \bibinfo {author} {\bibfnamefont {G.}~\bibnamefont {Long}},\ }\bibfield  {title} {\bibinfo {title} {A full quantum eigensolver for quantum chemistry simulations},\ }\href@noop {} {\bibfield  {journal} {\bibinfo  {journal} {Research}\ }\textbf {\bibinfo {volume} {2020}},\ \bibinfo {pages} {1486935} (\bibinfo {year} {2020})}\BibitemShut {NoStop}%
\bibitem [{\citenamefont {Fernández-Caramès}\ and\ \citenamefont {Fraga-Lamas}(2020)}]{Fernandez2020towards}%
  \BibitemOpen
  \bibfield  {author} {\bibinfo {author} {\bibfnamefont {T.~M.}\ \bibnamefont {Fernández-Caramès}}\ and\ \bibinfo {author} {\bibfnamefont {P.}~\bibnamefont {Fraga-Lamas}},\ }\bibfield  {title} {\bibinfo {title} {Towards post-quantum blockchain: A review on blockchain cryptography resistant to quantum computing attacks},\ }\href@noop {} {\bibfield  {journal} {\bibinfo  {journal} {IEEE Access}\ }\textbf {\bibinfo {volume} {8}},\ \bibinfo {pages} {21091} (\bibinfo {year} {2020})}\BibitemShut {NoStop}%
\bibitem [{\citenamefont {Zhou}\ \emph {et~al.}(2022)\citenamefont {Zhou}, \citenamefont {Cao}, \citenamefont {Lu}, \citenamefont {Wang}, \citenamefont {Bao}, \citenamefont {Jia}, \citenamefont {Fu}, \citenamefont {Yin},\ and\ \citenamefont {Chen}}]{zhou2022experimental}%
  \BibitemOpen
  \bibfield  {author} {\bibinfo {author} {\bibfnamefont {M.-G.}\ \bibnamefont {Zhou}}, \bibinfo {author} {\bibfnamefont {X.-Y.}\ \bibnamefont {Cao}}, \bibinfo {author} {\bibfnamefont {Y.-S.}\ \bibnamefont {Lu}}, \bibinfo {author} {\bibfnamefont {Y.}~\bibnamefont {Wang}}, \bibinfo {author} {\bibfnamefont {Y.}~\bibnamefont {Bao}}, \bibinfo {author} {\bibfnamefont {Z.-Y.}\ \bibnamefont {Jia}}, \bibinfo {author} {\bibfnamefont {Y.}~\bibnamefont {Fu}}, \bibinfo {author} {\bibfnamefont {H.-L.}\ \bibnamefont {Yin}},\ and\ \bibinfo {author} {\bibfnamefont {Z.-B.}\ \bibnamefont {Chen}},\ }\bibfield  {title} {\bibinfo {title} {Experimental quantum advantage with quantum coupon collector},\ }\href@noop {} {\bibfield  {journal} {\bibinfo  {journal} {Research}\ }\textbf {\bibinfo {volume} {2022}},\ \bibinfo {pages} {9798679} (\bibinfo {year} {2022})}\BibitemShut {NoStop}%
\bibitem [{\citenamefont {Huang}\ \emph {et~al.}(2022)\citenamefont {Huang}, \citenamefont {Yin}, \citenamefont {Chen},\ and\ \citenamefont {Wu}}]{huang2022quantum}%
  \BibitemOpen
  \bibfield  {author} {\bibinfo {author} {\bibfnamefont {S.}~\bibnamefont {Huang}}, \bibinfo {author} {\bibfnamefont {H.-L.}\ \bibnamefont {Yin}}, \bibinfo {author} {\bibfnamefont {Z.-B.}\ \bibnamefont {Chen}},\ and\ \bibinfo {author} {\bibfnamefont {S.}~\bibnamefont {Wu}},\ }\bibfield  {title} {\bibinfo {title} {Quantum-accelerated algorithms for generating random primitive polynomials over finite fields},\ }\href@noop {} {\bibfield  {journal} {\bibinfo  {journal} {arXiv preprint arXiv:2203.12884}\ } (\bibinfo {year} {2022})}\BibitemShut {NoStop}%
\bibitem [{\citenamefont {Pan}\ \emph {et~al.}(2021)\citenamefont {Pan}, \citenamefont {Chen}, \citenamefont {Sun},\ and\ \citenamefont {Zhang}}]{pan2021electric}%
  \BibitemOpen
  \bibfield  {author} {\bibinfo {author} {\bibfnamefont {N.}~\bibnamefont {Pan}}, \bibinfo {author} {\bibfnamefont {T.}~\bibnamefont {Chen}}, \bibinfo {author} {\bibfnamefont {H.}~\bibnamefont {Sun}},\ and\ \bibinfo {author} {\bibfnamefont {X.}~\bibnamefont {Zhang}},\ }\bibfield  {title} {\bibinfo {title} {Electric-circuit realization of fast quantum search},\ }\href@noop {} {\bibfield  {journal} {\bibinfo  {journal} {Research}\ }\textbf {\bibinfo {volume} {2021}},\ \bibinfo {pages} {9793071} (\bibinfo {year} {2021})}\BibitemShut {NoStop}%
\bibitem [{\citenamefont {Long}(2022)}]{long2022toward}%
  \BibitemOpen
  \bibfield  {author} {\bibinfo {author} {\bibfnamefont {G.-L.}\ \bibnamefont {Long}},\ }\bibfield  {title} {\bibinfo {title} {Toward applications of cloud quantum computation},\ }\href@noop {} {\bibfield  {journal} {\bibinfo  {journal} {Science China Physics, Mechanics \& Astronomy}\ }\textbf {\bibinfo {volume} {65}},\ \bibinfo {pages} {110361} (\bibinfo {year} {2022})}\BibitemShut {NoStop}%
\bibitem [{\citenamefont {Fitzi}\ \emph {et~al.}(2001{\natexlab{b}})\citenamefont {Fitzi}, \citenamefont {Gisin},\ and\ \citenamefont {Maurer}}]{Fitzi2001quantum}%
  \BibitemOpen
  \bibfield  {author} {\bibinfo {author} {\bibfnamefont {M.}~\bibnamefont {Fitzi}}, \bibinfo {author} {\bibfnamefont {N.}~\bibnamefont {Gisin}},\ and\ \bibinfo {author} {\bibfnamefont {U.}~\bibnamefont {Maurer}},\ }\bibfield  {title} {\bibinfo {title} {Quantum solution to the {B}yzantine agreement problem},\ }\href@noop {} {\bibfield  {journal} {\bibinfo  {journal} {Phys. Rev. Lett.}\ }\textbf {\bibinfo {volume} {87}},\ \bibinfo {pages} {217901} (\bibinfo {year} {2001}{\natexlab{b}})}\BibitemShut {NoStop}%
\bibitem [{\citenamefont {Gaertner}\ \emph {et~al.}(2008{\natexlab{a}})\citenamefont {Gaertner}, \citenamefont {Bourennane}, \citenamefont {Kurtsiefer}, \citenamefont {Cabello},\ and\ \citenamefont {Weinfurter}}]{gaertner2008xperimental}%
  \BibitemOpen
  \bibfield  {author} {\bibinfo {author} {\bibfnamefont {S.}~\bibnamefont {Gaertner}}, \bibinfo {author} {\bibfnamefont {M.}~\bibnamefont {Bourennane}}, \bibinfo {author} {\bibfnamefont {C.}~\bibnamefont {Kurtsiefer}}, \bibinfo {author} {\bibfnamefont {A.}~\bibnamefont {Cabello}},\ and\ \bibinfo {author} {\bibfnamefont {H.}~\bibnamefont {Weinfurter}},\ }\bibfield  {title} {\bibinfo {title} {Experimental demonstration of a quantum protocol for {B}yzantine agreement and liar detection},\ }\href@noop {} {\bibfield  {journal} {\bibinfo  {journal} {Phys. Rev. Lett.}\ }\textbf {\bibinfo {volume} {100}},\ \bibinfo {pages} {070504} (\bibinfo {year} {2008}{\natexlab{a}})}\BibitemShut {NoStop}%
\bibitem [{\citenamefont {Fitzi}\ \emph {et~al.}(2002)\citenamefont {Fitzi}, \citenamefont {Gottesman}, \citenamefont {Hirt}, \citenamefont {Holenstein},\ and\ \citenamefont {Smith}}]{fitzi2002detectable}%
  \BibitemOpen
  \bibfield  {author} {\bibinfo {author} {\bibfnamefont {M.}~\bibnamefont {Fitzi}}, \bibinfo {author} {\bibfnamefont {D.}~\bibnamefont {Gottesman}}, \bibinfo {author} {\bibfnamefont {M.}~\bibnamefont {Hirt}}, \bibinfo {author} {\bibfnamefont {T.}~\bibnamefont {Holenstein}},\ and\ \bibinfo {author} {\bibfnamefont {A.}~\bibnamefont {Smith}},\ }\bibfield  {title} {\bibinfo {title} {Detectable {B}yzantine agreement secure against faulty majorities},\ }in\ \href@noop {} {\emph {\bibinfo {booktitle} {Proceedings of the twenty-first annual symposium on Principles of distributed computing}}}\ (\bibinfo {year} {2002})\ pp.\ \bibinfo {pages} {118--126}\BibitemShut {NoStop}%
\bibitem [{\citenamefont {Iblisdir}\ and\ \citenamefont {Gisin}(2004)}]{Iblisdir2004byzantine}%
  \BibitemOpen
  \bibfield  {author} {\bibinfo {author} {\bibfnamefont {S.}~\bibnamefont {Iblisdir}}\ and\ \bibinfo {author} {\bibfnamefont {N.}~\bibnamefont {Gisin}},\ }\bibfield  {title} {\bibinfo {title} {Byzantine agreement with two quantum-key-distribution setups},\ }\href@noop {} {\bibfield  {journal} {\bibinfo  {journal} {Phys. Rev. A}\ }\textbf {\bibinfo {volume} {70}},\ \bibinfo {pages} {034306} (\bibinfo {year} {2004})}\BibitemShut {NoStop}%
\bibitem [{\citenamefont {Neigovzen}\ \emph {et~al.}(2008)\citenamefont {Neigovzen}, \citenamefont {Rod\'o}, \citenamefont {Adesso},\ and\ \citenamefont {Sanpera}}]{Neigovzen2008Multipartite}%
  \BibitemOpen
  \bibfield  {author} {\bibinfo {author} {\bibfnamefont {R.}~\bibnamefont {Neigovzen}}, \bibinfo {author} {\bibfnamefont {C.}~\bibnamefont {Rod\'o}}, \bibinfo {author} {\bibfnamefont {G.}~\bibnamefont {Adesso}},\ and\ \bibinfo {author} {\bibfnamefont {A.}~\bibnamefont {Sanpera}},\ }\bibfield  {title} {\bibinfo {title} {Multipartite continuous-variable solution for the {B}yzantine agreement problem},\ }\href@noop {} {\bibfield  {journal} {\bibinfo  {journal} {Phys. Rev. A}\ }\textbf {\bibinfo {volume} {77}},\ \bibinfo {pages} {062307} (\bibinfo {year} {2008})}\BibitemShut {NoStop}%
\bibitem [{\citenamefont {Rahaman}\ \emph {et~al.}(2015)\citenamefont {Rahaman}, \citenamefont {Wie\ifmmode~\acute{s}\else \'{s}\fi{}niak},\ and\ \citenamefont {\ifmmode~\dot{Z}\else \.{Z}\fi{}ukowski}}]{Rahaman2015Quantum}%
  \BibitemOpen
  \bibfield  {author} {\bibinfo {author} {\bibfnamefont {R.}~\bibnamefont {Rahaman}}, \bibinfo {author} {\bibfnamefont {M.}~\bibnamefont {Wie\ifmmode~\acute{s}\else \'{s}\fi{}niak}},\ and\ \bibinfo {author} {\bibfnamefont {M.}~\bibnamefont {\ifmmode~\dot{Z}\else \.{Z}\fi{}ukowski}},\ }\bibfield  {title} {\bibinfo {title} {Quantum {B}yzantine agreement via {H}ardy correlations and entanglement swapping},\ }\href@noop {} {\bibfield  {journal} {\bibinfo  {journal} {Phys. Rev. A}\ }\textbf {\bibinfo {volume} {92}},\ \bibinfo {pages} {042302} (\bibinfo {year} {2015})}\BibitemShut {NoStop}%
\bibitem [{\citenamefont {Smania}\ \emph {et~al.}(2016)\citenamefont {Smania}, \citenamefont {Elhassan}, \citenamefont {Tavakoli},\ and\ \citenamefont {Bourennane}}]{smania2016experimental}%
  \BibitemOpen
  \bibfield  {author} {\bibinfo {author} {\bibfnamefont {M.}~\bibnamefont {Smania}}, \bibinfo {author} {\bibfnamefont {A.~M.}\ \bibnamefont {Elhassan}}, \bibinfo {author} {\bibfnamefont {A.}~\bibnamefont {Tavakoli}},\ and\ \bibinfo {author} {\bibfnamefont {M.}~\bibnamefont {Bourennane}},\ }\bibfield  {title} {\bibinfo {title} {Experimental quantum multiparty communication protocols},\ }\href@noop {} {\bibfield  {journal} {\bibinfo  {journal} {npj Quantum Inf.}\ }\textbf {\bibinfo {volume} {2}},\ \bibinfo {pages} {16010} (\bibinfo {year} {2016})}\BibitemShut {NoStop}%
\bibitem [{\citenamefont {Ben-Or}\ and\ \citenamefont {Hassidim}(2005)}]{Ben2005fast}%
  \BibitemOpen
  \bibfield  {author} {\bibinfo {author} {\bibfnamefont {M.}~\bibnamefont {Ben-Or}}\ and\ \bibinfo {author} {\bibfnamefont {A.}~\bibnamefont {Hassidim}},\ }\bibfield  {title} {\bibinfo {title} {Fast quantum {B}yzantine agreement},\ }in\ \href@noop {} {\emph {\bibinfo {booktitle} {Proceedings of the Thirty-Seventh Annual ACM Symposium on Theory of Computing}}}\ (\bibinfo  {publisher} {Association for Computing Machinery},\ \bibinfo {address} {New York, NY, USA},\ \bibinfo {year} {2005})\ p.\ \bibinfo {pages} {481–485}\BibitemShut {NoStop}%
\bibitem [{\citenamefont {Taherkhani}\ \emph {et~al.}(2018)\citenamefont {Taherkhani}, \citenamefont {Navi},\ and\ \citenamefont {Van~Meter}}]{taherkhani2018resource}%
  \BibitemOpen
  \bibfield  {author} {\bibinfo {author} {\bibfnamefont {M.~A.}\ \bibnamefont {Taherkhani}}, \bibinfo {author} {\bibfnamefont {K.}~\bibnamefont {Navi}},\ and\ \bibinfo {author} {\bibfnamefont {R.}~\bibnamefont {Van~Meter}},\ }\bibfield  {title} {\bibinfo {title} {Resource-aware system architecture model for implementation of quantum aided {B}yzantine agreement on quantum repeater networks},\ }\href@noop {} {\bibfield  {journal} {\bibinfo  {journal} {Quantum Sci. Technol.}\ }\textbf {\bibinfo {volume} {3}},\ \bibinfo {pages} {014011} (\bibinfo {year} {2018})}\BibitemShut {NoStop}%
\bibitem [{\citenamefont {Sun}\ \emph {et~al.}(2020)\citenamefont {Sun}, \citenamefont {Kulicki},\ and\ \citenamefont {Sopek}}]{sun2020multi}%
  \BibitemOpen
  \bibfield  {author} {\bibinfo {author} {\bibfnamefont {X.}~\bibnamefont {Sun}}, \bibinfo {author} {\bibfnamefont {P.}~\bibnamefont {Kulicki}},\ and\ \bibinfo {author} {\bibfnamefont {M.}~\bibnamefont {Sopek}},\ }\bibfield  {title} {\bibinfo {title} {Multi-party quantum {B}yzantine agreement without entanglement},\ }\href@noop {} {\bibfield  {journal} {\bibinfo  {journal} {Entropy}\ }\textbf {\bibinfo {volume} {22}},\ \bibinfo {pages} {1152} (\bibinfo {year} {2020})}\BibitemShut {NoStop}%
\bibitem [{\citenamefont {Wang}\ \emph {et~al.}(2022)\citenamefont {Wang}, \citenamefont {Yu},\ and\ \citenamefont {Du}}]{wang2022quantum}%
  \BibitemOpen
  \bibfield  {author} {\bibinfo {author} {\bibfnamefont {W.}~\bibnamefont {Wang}}, \bibinfo {author} {\bibfnamefont {Y.}~\bibnamefont {Yu}},\ and\ \bibinfo {author} {\bibfnamefont {L.}~\bibnamefont {Du}},\ }\bibfield  {title} {\bibinfo {title} {Quantum blockchain based on asymmetric quantum encryption and a stake vote consensus algorithm},\ }\href@noop {} {\bibfield  {journal} {\bibinfo  {journal} {Sci. Rep.}\ }\textbf {\bibinfo {volume} {12}},\ \bibinfo {pages} {8606} (\bibinfo {year} {2022})}\BibitemShut {NoStop}%
\bibitem [{\citenamefont {Gao}\ \emph {et~al.}(2008)\citenamefont {Gao}, \citenamefont {Guo}, \citenamefont {Wen},\ and\ \citenamefont {Zhu}}]{Gao:2008:Common}%
  \BibitemOpen
  \bibfield  {author} {\bibinfo {author} {\bibfnamefont {F.}~\bibnamefont {Gao}}, \bibinfo {author} {\bibfnamefont {F.-Z.}\ \bibnamefont {Guo}}, \bibinfo {author} {\bibfnamefont {Q.-Y.}\ \bibnamefont {Wen}},\ and\ \bibinfo {author} {\bibfnamefont {F.-C.}\ \bibnamefont {Zhu}},\ }\bibfield  {title} {\bibinfo {title} {Comment on ``experimental demonstration of a quantum protocol for byzantine agreement and liar detection''},\ }\href@noop {} {\bibfield  {journal} {\bibinfo  {journal} {Phys. Rev. Lett.}\ }\textbf {\bibinfo {volume} {101}},\ \bibinfo {pages} {208901} (\bibinfo {year} {2008})}\BibitemShut {NoStop}%
\bibitem [{\citenamefont {Gaertner}\ \emph {et~al.}(2008{\natexlab{b}})\citenamefont {Gaertner}, \citenamefont {Bourennane}, \citenamefont {Kurtsiefer}, \citenamefont {Cabello},\ and\ \citenamefont {Weinfurter}}]{Gaertner:2008:Reply}%
  \BibitemOpen
  \bibfield  {author} {\bibinfo {author} {\bibfnamefont {S.}~\bibnamefont {Gaertner}}, \bibinfo {author} {\bibfnamefont {M.}~\bibnamefont {Bourennane}}, \bibinfo {author} {\bibfnamefont {C.}~\bibnamefont {Kurtsiefer}}, \bibinfo {author} {\bibfnamefont {A.}~\bibnamefont {Cabello}},\ and\ \bibinfo {author} {\bibfnamefont {H.}~\bibnamefont {Weinfurter}},\ }\bibfield  {title} {\bibinfo {title} {Gaertner et al. reply:},\ }\href@noop {} {\bibfield  {journal} {\bibinfo  {journal} {Phys. Rev. Lett.}\ }\textbf {\bibinfo {volume} {101}},\ \bibinfo {pages} {208902} (\bibinfo {year} {2008}{\natexlab{b}})}\BibitemShut {NoStop}%
\bibitem [{\citenamefont {Kleinberg}\ and\ \citenamefont {Tardos}(2006)}]{kleinberg2006algorithm}%
  \BibitemOpen
  \bibfield  {author} {\bibinfo {author} {\bibfnamefont {J.}~\bibnamefont {Kleinberg}}\ and\ \bibinfo {author} {\bibfnamefont {E.}~\bibnamefont {Tardos}},\ }\href@noop {} {\emph {\bibinfo {title} {Algorithm design}}}\ (\bibinfo  {publisher} {Pearson Education India},\ \bibinfo {year} {2006})\BibitemShut {NoStop}%
\bibitem [{\citenamefont {Gottesman}\ and\ \citenamefont {Chuang}(2001)}]{gottesman2001quantum}%
  \BibitemOpen
  \bibfield  {author} {\bibinfo {author} {\bibfnamefont {D.}~\bibnamefont {Gottesman}}\ and\ \bibinfo {author} {\bibfnamefont {I.}~\bibnamefont {Chuang}},\ }\bibfield  {title} {\bibinfo {title} {Quantum digital signatures},\ }\href@noop {} {\bibfield  {journal} {\bibinfo  {journal} {arXiv preprint quant-ph/0105032}\ } (\bibinfo {year} {2001})}\BibitemShut {NoStop}%
\bibitem [{\citenamefont {Dunjko}\ \emph {et~al.}(2014)\citenamefont {Dunjko}, \citenamefont {Wallden},\ and\ \citenamefont {Andersson}}]{dunjko2014quantum}%
  \BibitemOpen
  \bibfield  {author} {\bibinfo {author} {\bibfnamefont {V.}~\bibnamefont {Dunjko}}, \bibinfo {author} {\bibfnamefont {P.}~\bibnamefont {Wallden}},\ and\ \bibinfo {author} {\bibfnamefont {E.}~\bibnamefont {Andersson}},\ }\bibfield  {title} {\bibinfo {title} {Quantum digital signatures without quantum memory},\ }\href@noop {} {\bibfield  {journal} {\bibinfo  {journal} {Phys. Rev. Lett.}\ }\textbf {\bibinfo {volume} {112}},\ \bibinfo {pages} {040502} (\bibinfo {year} {2014})}\BibitemShut {NoStop}%
\bibitem [{\citenamefont {Roehsner}\ \emph {et~al.}(2018)\citenamefont {Roehsner}, \citenamefont {Kettlewell}, \citenamefont {Batalh{\~a}o}, \citenamefont {Fitzsimons},\ and\ \citenamefont {Walther}}]{roehsner2018quantum}%
  \BibitemOpen
  \bibfield  {author} {\bibinfo {author} {\bibfnamefont {M.-C.}\ \bibnamefont {Roehsner}}, \bibinfo {author} {\bibfnamefont {J.~A.}\ \bibnamefont {Kettlewell}}, \bibinfo {author} {\bibfnamefont {T.~B.}\ \bibnamefont {Batalh{\~a}o}}, \bibinfo {author} {\bibfnamefont {J.~F.}\ \bibnamefont {Fitzsimons}},\ and\ \bibinfo {author} {\bibfnamefont {P.}~\bibnamefont {Walther}},\ }\bibfield  {title} {\bibinfo {title} {Quantum advantage for probabilistic one-time programs},\ }\href@noop {} {\bibfield  {journal} {\bibinfo  {journal} {Nature Commun.}\ }\textbf {\bibinfo {volume} {9}},\ \bibinfo {pages} {5225} (\bibinfo {year} {2018})}\BibitemShut {NoStop}%
\bibitem [{\citenamefont {Amiri}\ \emph {et~al.}(2016)\citenamefont {Amiri}, \citenamefont {Wallden}, \citenamefont {Kent},\ and\ \citenamefont {Andersson}}]{amiri2016secure}%
  \BibitemOpen
  \bibfield  {author} {\bibinfo {author} {\bibfnamefont {R.}~\bibnamefont {Amiri}}, \bibinfo {author} {\bibfnamefont {P.}~\bibnamefont {Wallden}}, \bibinfo {author} {\bibfnamefont {A.}~\bibnamefont {Kent}},\ and\ \bibinfo {author} {\bibfnamefont {E.}~\bibnamefont {Andersson}},\ }\bibfield  {title} {\bibinfo {title} {Secure quantum signatures using insecure quantum channels},\ }\href@noop {} {\bibfield  {journal} {\bibinfo  {journal} {Phys. Rev. A}\ }\textbf {\bibinfo {volume} {93}},\ \bibinfo {pages} {032325} (\bibinfo {year} {2016})}\BibitemShut {NoStop}%
\bibitem [{\citenamefont {Puthoor}\ \emph {et~al.}(2016)\citenamefont {Puthoor}, \citenamefont {Amiri}, \citenamefont {Wallden}, \citenamefont {Curty},\ and\ \citenamefont {Andersson}}]{Puthoor2016Mea}%
  \BibitemOpen
  \bibfield  {author} {\bibinfo {author} {\bibfnamefont {I.~V.}\ \bibnamefont {Puthoor}}, \bibinfo {author} {\bibfnamefont {R.}~\bibnamefont {Amiri}}, \bibinfo {author} {\bibfnamefont {P.}~\bibnamefont {Wallden}}, \bibinfo {author} {\bibfnamefont {M.}~\bibnamefont {Curty}},\ and\ \bibinfo {author} {\bibfnamefont {E.}~\bibnamefont {Andersson}},\ }\bibfield  {title} {\bibinfo {title} {Measurement-device-independent quantum digital signatures},\ }\href@noop {} {\bibfield  {journal} {\bibinfo  {journal} {Phys. Rev. A}\ }\textbf {\bibinfo {volume} {94}},\ \bibinfo {pages} {022328} (\bibinfo {year} {2016})}\BibitemShut {NoStop}%
\bibitem [{\citenamefont {Roberts}\ \emph {et~al.}(2017)\citenamefont {Roberts}, \citenamefont {Lucamarini}, \citenamefont {Yuan}, \citenamefont {Dynes}, \citenamefont {Comandar}, \citenamefont {Sharpe}, \citenamefont {Shields}, \citenamefont {Curty}, \citenamefont {Puthoor},\ and\ \citenamefont {Andersson}}]{roberts2017experimental}%
  \BibitemOpen
  \bibfield  {author} {\bibinfo {author} {\bibfnamefont {G.}~\bibnamefont {Roberts}}, \bibinfo {author} {\bibfnamefont {M.}~\bibnamefont {Lucamarini}}, \bibinfo {author} {\bibfnamefont {Z.}~\bibnamefont {Yuan}}, \bibinfo {author} {\bibfnamefont {J.}~\bibnamefont {Dynes}}, \bibinfo {author} {\bibfnamefont {L.}~\bibnamefont {Comandar}}, \bibinfo {author} {\bibfnamefont {A.}~\bibnamefont {Sharpe}}, \bibinfo {author} {\bibfnamefont {A.}~\bibnamefont {Shields}}, \bibinfo {author} {\bibfnamefont {M.}~\bibnamefont {Curty}}, \bibinfo {author} {\bibfnamefont {I.}~\bibnamefont {Puthoor}},\ and\ \bibinfo {author} {\bibfnamefont {E.}~\bibnamefont {Andersson}},\ }\bibfield  {title} {\bibinfo {title} {Experimental measurement-device-independent quantum digital signatures},\ }\href@noop {} {\bibfield  {journal} {\bibinfo  {journal} {Nat. Commun.}\ }\textbf {\bibinfo {volume} {8}},\ \bibinfo {pages} {1098} (\bibinfo {year} {2017})}\BibitemShut {NoStop}%
\bibitem [{\citenamefont {Collins}\ \emph {et~al.}(2017)\citenamefont {Collins}, \citenamefont {Amiri}, \citenamefont {Fujiwara}, \citenamefont {Honjo}, \citenamefont {Shimizu}, \citenamefont {Tamaki}, \citenamefont {Takeoka}, \citenamefont {Sasaki}, \citenamefont {Andersson},\ and\ \citenamefont {Buller}}]{collins2017experimental}%
  \BibitemOpen
  \bibfield  {author} {\bibinfo {author} {\bibfnamefont {R.~J.}\ \bibnamefont {Collins}}, \bibinfo {author} {\bibfnamefont {R.}~\bibnamefont {Amiri}}, \bibinfo {author} {\bibfnamefont {M.}~\bibnamefont {Fujiwara}}, \bibinfo {author} {\bibfnamefont {T.}~\bibnamefont {Honjo}}, \bibinfo {author} {\bibfnamefont {K.}~\bibnamefont {Shimizu}}, \bibinfo {author} {\bibfnamefont {K.}~\bibnamefont {Tamaki}}, \bibinfo {author} {\bibfnamefont {M.}~\bibnamefont {Takeoka}}, \bibinfo {author} {\bibfnamefont {M.}~\bibnamefont {Sasaki}}, \bibinfo {author} {\bibfnamefont {E.}~\bibnamefont {Andersson}},\ and\ \bibinfo {author} {\bibfnamefont {G.~S.}\ \bibnamefont {Buller}},\ }\bibfield  {title} {\bibinfo {title} {Experimental demonstration of quantum digital signatures over 43 d{B} channel loss using differential phase shift quantum key distribution},\ }\href@noop {} {\bibfield  {journal} {\bibinfo  {journal} {Sci. Rep.}\ }\textbf {\bibinfo {volume} {7}},\ \bibinfo {pages} {3235} (\bibinfo {year} {2017})}\BibitemShut {NoStop}%
\bibitem [{\citenamefont {An}\ \emph {et~al.}(2019)\citenamefont {An}, \citenamefont {Zhang}, \citenamefont {Zhang}, \citenamefont {Chen}, \citenamefont {Wang}, \citenamefont {Yin}, \citenamefont {Wang}, \citenamefont {He}, \citenamefont {Hao}, \citenamefont {Liu} \emph {et~al.}}]{an2019practical}%
  \BibitemOpen
  \bibfield  {author} {\bibinfo {author} {\bibfnamefont {X.-B.}\ \bibnamefont {An}}, \bibinfo {author} {\bibfnamefont {H.}~\bibnamefont {Zhang}}, \bibinfo {author} {\bibfnamefont {C.-M.}\ \bibnamefont {Zhang}}, \bibinfo {author} {\bibfnamefont {W.}~\bibnamefont {Chen}}, \bibinfo {author} {\bibfnamefont {S.}~\bibnamefont {Wang}}, \bibinfo {author} {\bibfnamefont {Z.-Q.}\ \bibnamefont {Yin}}, \bibinfo {author} {\bibfnamefont {Q.}~\bibnamefont {Wang}}, \bibinfo {author} {\bibfnamefont {D.-Y.}\ \bibnamefont {He}}, \bibinfo {author} {\bibfnamefont {P.-L.}\ \bibnamefont {Hao}}, \bibinfo {author} {\bibfnamefont {S.-F.}\ \bibnamefont {Liu}}, \emph {et~al.},\ }\bibfield  {title} {\bibinfo {title} {Practical quantum digital signature with a gigahertz {BB}84 quantum key distribution system},\ }\href@noop {} {\bibfield  {journal} {\bibinfo  {journal} {Opt. Lett.}\ }\textbf {\bibinfo {volume} {44}},\ \bibinfo {pages} {139} (\bibinfo {year} {2019})}\BibitemShut {NoStop}%
\bibitem [{\citenamefont {Thornton}\ \emph {et~al.}(2019)\citenamefont {Thornton}, \citenamefont {Scott}, \citenamefont {Croal},\ and\ \citenamefont {Korolkova}}]{thornton2019continuous}%
  \BibitemOpen
  \bibfield  {author} {\bibinfo {author} {\bibfnamefont {M.}~\bibnamefont {Thornton}}, \bibinfo {author} {\bibfnamefont {H.}~\bibnamefont {Scott}}, \bibinfo {author} {\bibfnamefont {C.}~\bibnamefont {Croal}},\ and\ \bibinfo {author} {\bibfnamefont {N.}~\bibnamefont {Korolkova}},\ }\bibfield  {title} {\bibinfo {title} {Continuous-variable quantum digital signatures over insecure channels},\ }\href@noop {} {\bibfield  {journal} {\bibinfo  {journal} {Phys. Rev. A}\ }\textbf {\bibinfo {volume} {99}},\ \bibinfo {pages} {032341} (\bibinfo {year} {2019})}\BibitemShut {NoStop}%
\bibitem [{\citenamefont {Richter}\ \emph {et~al.}(2021)\citenamefont {Richter}, \citenamefont {Thornton}, \citenamefont {Khan}, \citenamefont {Scott}, \citenamefont {Jaksch}, \citenamefont {Vogl}, \citenamefont {Stiller}, \citenamefont {Leuchs}, \citenamefont {Marquardt},\ and\ \citenamefont {Korolkova}}]{richter2021agile}%
  \BibitemOpen
  \bibfield  {author} {\bibinfo {author} {\bibfnamefont {S.}~\bibnamefont {Richter}}, \bibinfo {author} {\bibfnamefont {M.}~\bibnamefont {Thornton}}, \bibinfo {author} {\bibfnamefont {I.}~\bibnamefont {Khan}}, \bibinfo {author} {\bibfnamefont {H.}~\bibnamefont {Scott}}, \bibinfo {author} {\bibfnamefont {K.}~\bibnamefont {Jaksch}}, \bibinfo {author} {\bibfnamefont {U.}~\bibnamefont {Vogl}}, \bibinfo {author} {\bibfnamefont {B.}~\bibnamefont {Stiller}}, \bibinfo {author} {\bibfnamefont {G.}~\bibnamefont {Leuchs}}, \bibinfo {author} {\bibfnamefont {C.}~\bibnamefont {Marquardt}},\ and\ \bibinfo {author} {\bibfnamefont {N.}~\bibnamefont {Korolkova}},\ }\bibfield  {title} {\bibinfo {title} {Agile and versatile quantum communication: Signatures and secrets},\ }\href@noop {} {\bibfield  {journal} {\bibinfo  {journal} {Phys. Rev. X}\ }\textbf {\bibinfo {volume} {11}},\ \bibinfo {pages} {011038} (\bibinfo {year} {2021})}\BibitemShut {NoStop}%
\bibitem [{\citenamefont {Qin}\ \emph {et~al.}(2022)\citenamefont {Qin}, \citenamefont {Jiang}, \citenamefont {Yu},\ and\ \citenamefont {Wang}}]{qin2022quantum}%
  \BibitemOpen
  \bibfield  {author} {\bibinfo {author} {\bibfnamefont {J.-Q.}\ \bibnamefont {Qin}}, \bibinfo {author} {\bibfnamefont {C.}~\bibnamefont {Jiang}}, \bibinfo {author} {\bibfnamefont {Y.-L.}\ \bibnamefont {Yu}},\ and\ \bibinfo {author} {\bibfnamefont {X.-B.}\ \bibnamefont {Wang}},\ }\bibfield  {title} {\bibinfo {title} {Quantum digital signatures with random pairing},\ }\href@noop {} {\bibfield  {journal} {\bibinfo  {journal} {Phys. Rev. Applied}\ }\textbf {\bibinfo {volume} {17}},\ \bibinfo {pages} {044047} (\bibinfo {year} {2022})}\BibitemShut {NoStop}%
\bibitem [{\citenamefont {Yin}\ \emph {et~al.}(2017{\natexlab{a}})\citenamefont {Yin}, \citenamefont {Wang}, \citenamefont {Tang}, \citenamefont {Zhao}, \citenamefont {Liu}, \citenamefont {Sun}, \citenamefont {Zhang}, \citenamefont {Li}, \citenamefont {Puthoor}, \citenamefont {You} \emph {et~al.}}]{yin2017experimental}%
  \BibitemOpen
  \bibfield  {author} {\bibinfo {author} {\bibfnamefont {H.-L.}\ \bibnamefont {Yin}}, \bibinfo {author} {\bibfnamefont {W.-L.}\ \bibnamefont {Wang}}, \bibinfo {author} {\bibfnamefont {Y.-L.}\ \bibnamefont {Tang}}, \bibinfo {author} {\bibfnamefont {Q.}~\bibnamefont {Zhao}}, \bibinfo {author} {\bibfnamefont {H.}~\bibnamefont {Liu}}, \bibinfo {author} {\bibfnamefont {X.-X.}\ \bibnamefont {Sun}}, \bibinfo {author} {\bibfnamefont {W.-J.}\ \bibnamefont {Zhang}}, \bibinfo {author} {\bibfnamefont {H.}~\bibnamefont {Li}}, \bibinfo {author} {\bibfnamefont {I.~V.}\ \bibnamefont {Puthoor}}, \bibinfo {author} {\bibfnamefont {L.-X.}\ \bibnamefont {You}}, \emph {et~al.},\ }\bibfield  {title} {\bibinfo {title} {Experimental measurement-device-independent quantum digital signatures over a metropolitan network},\ }\href@noop {} {\bibfield  {journal} {\bibinfo  {journal} {Phys. Rev. A}\ }\textbf {\bibinfo {volume} {95}},\ \bibinfo {pages} {042338} (\bibinfo {year} {2017}{\natexlab{a}})}\BibitemShut {NoStop}%
\bibitem [{\citenamefont {Yin}\ \emph {et~al.}(2016)\citenamefont {Yin}, \citenamefont {Fu},\ and\ \citenamefont {Chen}}]{yin2016practical}%
  \BibitemOpen
  \bibfield  {author} {\bibinfo {author} {\bibfnamefont {H.-L.}\ \bibnamefont {Yin}}, \bibinfo {author} {\bibfnamefont {Y.}~\bibnamefont {Fu}},\ and\ \bibinfo {author} {\bibfnamefont {Z.-B.}\ \bibnamefont {Chen}},\ }\bibfield  {title} {\bibinfo {title} {Practical quantum digital signature},\ }\href@noop {} {\bibfield  {journal} {\bibinfo  {journal} {Phys. Rev. A}\ }\textbf {\bibinfo {volume} {93}},\ \bibinfo {pages} {032316} (\bibinfo {year} {2016})}\BibitemShut {NoStop}%
\bibitem [{\citenamefont {Yin}\ \emph {et~al.}(2017{\natexlab{b}})\citenamefont {Yin}, \citenamefont {Fu}, \citenamefont {Liu}, \citenamefont {Tang}, \citenamefont {Wang}, \citenamefont {You}, \citenamefont {Zhang}, \citenamefont {Chen}, \citenamefont {Wang}, \citenamefont {Zhang} \emph {et~al.}}]{yin2017experiment}%
  \BibitemOpen
  \bibfield  {author} {\bibinfo {author} {\bibfnamefont {H.-L.}\ \bibnamefont {Yin}}, \bibinfo {author} {\bibfnamefont {Y.}~\bibnamefont {Fu}}, \bibinfo {author} {\bibfnamefont {H.}~\bibnamefont {Liu}}, \bibinfo {author} {\bibfnamefont {Q.-J.}\ \bibnamefont {Tang}}, \bibinfo {author} {\bibfnamefont {J.}~\bibnamefont {Wang}}, \bibinfo {author} {\bibfnamefont {L.-X.}\ \bibnamefont {You}}, \bibinfo {author} {\bibfnamefont {W.-J.}\ \bibnamefont {Zhang}}, \bibinfo {author} {\bibfnamefont {S.-J.}\ \bibnamefont {Chen}}, \bibinfo {author} {\bibfnamefont {Z.}~\bibnamefont {Wang}}, \bibinfo {author} {\bibfnamefont {Q.}~\bibnamefont {Zhang}}, \emph {et~al.},\ }\bibfield  {title} {\bibinfo {title} {Experimental quantum digital signature over 102 km},\ }\href@noop {} {\bibfield  {journal} {\bibinfo  {journal} {Phys. Rev. A}\ }\textbf {\bibinfo {volume} {95}},\ \bibinfo {pages} {032334} (\bibinfo {year} {2017}{\natexlab{b}})}\BibitemShut {NoStop}%
\bibitem [{\citenamefont {Lu}\ \emph {et~al.}(2021)\citenamefont {Lu}, \citenamefont {Cao}, \citenamefont {Weng}, \citenamefont {Gu}, \citenamefont {Xie}, \citenamefont {Zhou}, \citenamefont {Yin},\ and\ \citenamefont {Chen}}]{lu2021efficient}%
  \BibitemOpen
  \bibfield  {author} {\bibinfo {author} {\bibfnamefont {Y.-S.}\ \bibnamefont {Lu}}, \bibinfo {author} {\bibfnamefont {X.-Y.}\ \bibnamefont {Cao}}, \bibinfo {author} {\bibfnamefont {C.-X.}\ \bibnamefont {Weng}}, \bibinfo {author} {\bibfnamefont {J.}~\bibnamefont {Gu}}, \bibinfo {author} {\bibfnamefont {Y.-M.}\ \bibnamefont {Xie}}, \bibinfo {author} {\bibfnamefont {M.-G.}\ \bibnamefont {Zhou}}, \bibinfo {author} {\bibfnamefont {H.-L.}\ \bibnamefont {Yin}},\ and\ \bibinfo {author} {\bibfnamefont {Z.-B.}\ \bibnamefont {Chen}},\ }\bibfield  {title} {\bibinfo {title} {Efficient quantum digital signatures without symmetrization step},\ }\href@noop {} {\bibfield  {journal} {\bibinfo  {journal} {Opt. Express}\ }\textbf {\bibinfo {volume} {29}},\ \bibinfo {pages} {10162} (\bibinfo {year} {2021})}\BibitemShut {NoStop}%
\bibitem [{\citenamefont {Weng}\ \emph {et~al.}(2021)\citenamefont {Weng}, \citenamefont {Lu}, \citenamefont {Gao}, \citenamefont {Xie}, \citenamefont {Gu}, \citenamefont {Li}, \citenamefont {Li}, \citenamefont {Yin},\ and\ \citenamefont {Chen}}]{Weng2021secure}%
  \BibitemOpen
  \bibfield  {author} {\bibinfo {author} {\bibfnamefont {C.-X.}\ \bibnamefont {Weng}}, \bibinfo {author} {\bibfnamefont {Y.-S.}\ \bibnamefont {Lu}}, \bibinfo {author} {\bibfnamefont {R.-Q.}\ \bibnamefont {Gao}}, \bibinfo {author} {\bibfnamefont {Y.-M.}\ \bibnamefont {Xie}}, \bibinfo {author} {\bibfnamefont {J.}~\bibnamefont {Gu}}, \bibinfo {author} {\bibfnamefont {C.-L.}\ \bibnamefont {Li}}, \bibinfo {author} {\bibfnamefont {B.-H.}\ \bibnamefont {Li}}, \bibinfo {author} {\bibfnamefont {H.-L.}\ \bibnamefont {Yin}},\ and\ \bibinfo {author} {\bibfnamefont {Z.-B.}\ \bibnamefont {Chen}},\ }\bibfield  {title} {\bibinfo {title} {Secure and practical multiparty quantum digital signatures},\ }\href@noop {} {\bibfield  {journal} {\bibinfo  {journal} {Opt. Express}\ }\textbf {\bibinfo {volume} {29}},\ \bibinfo {pages} {27661} (\bibinfo {year} {2021})}\BibitemShut {NoStop}%
\bibitem [{\citenamefont {Yin}\ \emph {et~al.}(2023)\citenamefont {Yin}, \citenamefont {Fu}, \citenamefont {Li}, \citenamefont {Weng}, \citenamefont {Li}, \citenamefont {Gu}, \citenamefont {Lu}, \citenamefont {Huang},\ and\ \citenamefont {Chen}}]{yin2021experimental}%
  \BibitemOpen
  \bibfield  {author} {\bibinfo {author} {\bibfnamefont {H.-L.}\ \bibnamefont {Yin}}, \bibinfo {author} {\bibfnamefont {Y.}~\bibnamefont {Fu}}, \bibinfo {author} {\bibfnamefont {C.-L.}\ \bibnamefont {Li}}, \bibinfo {author} {\bibfnamefont {C.-X.}\ \bibnamefont {Weng}}, \bibinfo {author} {\bibfnamefont {B.-H.}\ \bibnamefont {Li}}, \bibinfo {author} {\bibfnamefont {J.}~\bibnamefont {Gu}}, \bibinfo {author} {\bibfnamefont {Y.-S.}\ \bibnamefont {Lu}}, \bibinfo {author} {\bibfnamefont {S.}~\bibnamefont {Huang}},\ and\ \bibinfo {author} {\bibfnamefont {Z.-B.}\ \bibnamefont {Chen}},\ }\bibfield  {title} {\bibinfo {title} {{Experimental quantum secure network with digital signatures and encryption}},\ }\href@noop {} {\bibfield  {journal} {\bibinfo  {journal} {Natl. Sci. Rev.}\ }\textbf {\bibinfo {volume} {10}},\ \bibinfo {pages} {nwac228} (\bibinfo {year} {2023})}\BibitemShut {NoStop}%
\bibitem [{\citenamefont {Li}\ \emph {et~al.}(2023)\citenamefont {Li}, \citenamefont {Xie}, \citenamefont {Cao}, \citenamefont {Li}, \citenamefont {Fu}, \citenamefont {Yin},\ and\ \citenamefont {Chen}}]{li2023one}%
  \BibitemOpen
  \bibfield  {author} {\bibinfo {author} {\bibfnamefont {B.-H.}\ \bibnamefont {Li}}, \bibinfo {author} {\bibfnamefont {Y.-M.}\ \bibnamefont {Xie}}, \bibinfo {author} {\bibfnamefont {X.-Y.}\ \bibnamefont {Cao}}, \bibinfo {author} {\bibfnamefont {C.-L.}\ \bibnamefont {Li}}, \bibinfo {author} {\bibfnamefont {Y.}~\bibnamefont {Fu}}, \bibinfo {author} {\bibfnamefont {H.-L.}\ \bibnamefont {Yin}},\ and\ \bibinfo {author} {\bibfnamefont {Z.-B.}\ \bibnamefont {Chen}},\ }\bibfield  {title} {\bibinfo {title} {One-time universal hashing quantum digital signatures without perfect keys},\ }\href@noop {} {\bibfield  {journal} {\bibinfo  {journal} {Phys. Rev. Appl.}\ }\textbf {\bibinfo {volume} {20}},\ \bibinfo {pages} {044011} (\bibinfo {year} {2023})}\BibitemShut {NoStop}%
\bibitem [{\citenamefont {Yin}\ \emph {et~al.}(2020)\citenamefont {Yin}, \citenamefont {Liu}, \citenamefont {Dai}, \citenamefont {Ci}, \citenamefont {Gu}, \citenamefont {Gao}, \citenamefont {Wang},\ and\ \citenamefont {Shen}}]{yin2020experimental}%
  \BibitemOpen
  \bibfield  {author} {\bibinfo {author} {\bibfnamefont {H.-L.}\ \bibnamefont {Yin}}, \bibinfo {author} {\bibfnamefont {P.}~\bibnamefont {Liu}}, \bibinfo {author} {\bibfnamefont {W.-W.}\ \bibnamefont {Dai}}, \bibinfo {author} {\bibfnamefont {Z.-H.}\ \bibnamefont {Ci}}, \bibinfo {author} {\bibfnamefont {J.}~\bibnamefont {Gu}}, \bibinfo {author} {\bibfnamefont {T.}~\bibnamefont {Gao}}, \bibinfo {author} {\bibfnamefont {Q.-W.}\ \bibnamefont {Wang}},\ and\ \bibinfo {author} {\bibfnamefont {Z.-Y.}\ \bibnamefont {Shen}},\ }\bibfield  {title} {\bibinfo {title} {Experimental composable security decoy-state quantum key distribution using time-phase encoding},\ }\href@noop {} {\bibfield  {journal} {\bibinfo  {journal} {Opt. Express}\ }\textbf {\bibinfo {volume} {28}},\ \bibinfo {pages} {29479} (\bibinfo {year} {2020})}\BibitemShut {NoStop}%
\bibitem [{\citenamefont {Xu}\ \emph {et~al.}(2020)\citenamefont {Xu}, \citenamefont {Ma}, \citenamefont {Zhang}, \citenamefont {Lo},\ and\ \citenamefont {Pan}}]{xu2020secure}%
  \BibitemOpen
  \bibfield  {author} {\bibinfo {author} {\bibfnamefont {F.}~\bibnamefont {Xu}}, \bibinfo {author} {\bibfnamefont {X.}~\bibnamefont {Ma}}, \bibinfo {author} {\bibfnamefont {Q.}~\bibnamefont {Zhang}}, \bibinfo {author} {\bibfnamefont {H.-K.}\ \bibnamefont {Lo}},\ and\ \bibinfo {author} {\bibfnamefont {J.-W.}\ \bibnamefont {Pan}},\ }\bibfield  {title} {\bibinfo {title} {Secure quantum key distribution with realistic devices},\ }\href@noop {} {\bibfield  {journal} {\bibinfo  {journal} {Rev. Mod. Phys.}\ }\textbf {\bibinfo {volume} {92}},\ \bibinfo {pages} {025002} (\bibinfo {year} {2020})}\BibitemShut {NoStop}%
\bibitem [{\citenamefont {Pirandola}\ \emph {et~al.}(2020)\citenamefont {Pirandola}, \citenamefont {Andersen}, \citenamefont {Banchi}, \citenamefont {Berta}, \citenamefont {Bunandar}, \citenamefont {Colbeck}, \citenamefont {Englund}, \citenamefont {Gehring}, \citenamefont {Lupo}, \citenamefont {Ottaviani} \emph {et~al.}}]{pirandola2020advances}%
  \BibitemOpen
  \bibfield  {author} {\bibinfo {author} {\bibfnamefont {S.}~\bibnamefont {Pirandola}}, \bibinfo {author} {\bibfnamefont {U.~L.}\ \bibnamefont {Andersen}}, \bibinfo {author} {\bibfnamefont {L.}~\bibnamefont {Banchi}}, \bibinfo {author} {\bibfnamefont {M.}~\bibnamefont {Berta}}, \bibinfo {author} {\bibfnamefont {D.}~\bibnamefont {Bunandar}}, \bibinfo {author} {\bibfnamefont {R.}~\bibnamefont {Colbeck}}, \bibinfo {author} {\bibfnamefont {D.}~\bibnamefont {Englund}}, \bibinfo {author} {\bibfnamefont {T.}~\bibnamefont {Gehring}}, \bibinfo {author} {\bibfnamefont {C.}~\bibnamefont {Lupo}}, \bibinfo {author} {\bibfnamefont {C.}~\bibnamefont {Ottaviani}}, \emph {et~al.},\ }\bibfield  {title} {\bibinfo {title} {Advances in quantum cryptography},\ }\href@noop {} {\bibfield  {journal} {\bibinfo  {journal} {Adv. Opt. Photon.}\ }\textbf {\bibinfo {volume} {12}},\ \bibinfo {pages} {1012} (\bibinfo {year} {2020})}\BibitemShut {NoStop}%
\bibitem [{\citenamefont {Liu}\ \emph {et~al.}(2021)\citenamefont {Liu}, \citenamefont {Li}, \citenamefont {Xie}, \citenamefont {Weng}, \citenamefont {Gu}, \citenamefont {Cao}, \citenamefont {Lu}, \citenamefont {Li}, \citenamefont {Yin},\ and\ \citenamefont {Chen}}]{liu2021homodyne}%
  \BibitemOpen
  \bibfield  {author} {\bibinfo {author} {\bibfnamefont {W.-B.}\ \bibnamefont {Liu}}, \bibinfo {author} {\bibfnamefont {C.-L.}\ \bibnamefont {Li}}, \bibinfo {author} {\bibfnamefont {Y.-M.}\ \bibnamefont {Xie}}, \bibinfo {author} {\bibfnamefont {C.-X.}\ \bibnamefont {Weng}}, \bibinfo {author} {\bibfnamefont {J.}~\bibnamefont {Gu}}, \bibinfo {author} {\bibfnamefont {X.-Y.}\ \bibnamefont {Cao}}, \bibinfo {author} {\bibfnamefont {Y.-S.}\ \bibnamefont {Lu}}, \bibinfo {author} {\bibfnamefont {B.-H.}\ \bibnamefont {Li}}, \bibinfo {author} {\bibfnamefont {H.-L.}\ \bibnamefont {Yin}},\ and\ \bibinfo {author} {\bibfnamefont {Z.-B.}\ \bibnamefont {Chen}},\ }\bibfield  {title} {\bibinfo {title} {Homodyne detection quadrature phase shift keying continuous-variable quantum key distribution with high excess noise tolerance},\ }\href@noop {} {\bibfield  {journal} {\bibinfo  {journal} {PRX Quantum}\ }\textbf {\bibinfo {volume} {2}},\ \bibinfo {pages} {040334} (\bibinfo {year} {2021})}\BibitemShut {NoStop}%
\bibitem [{\citenamefont {Lo}\ \emph {et~al.}(2012)\citenamefont {Lo}, \citenamefont {Curty},\ and\ \citenamefont {Qi}}]{lo2012measurement}%
  \BibitemOpen
  \bibfield  {author} {\bibinfo {author} {\bibfnamefont {H.-K.}\ \bibnamefont {Lo}}, \bibinfo {author} {\bibfnamefont {M.}~\bibnamefont {Curty}},\ and\ \bibinfo {author} {\bibfnamefont {B.}~\bibnamefont {Qi}},\ }\bibfield  {title} {\bibinfo {title} {Measurement-device-independent quantum key distribution},\ }\href@noop {} {\bibfield  {journal} {\bibinfo  {journal} {Phys. Rev. Lett.}\ }\textbf {\bibinfo {volume} {108}},\ \bibinfo {pages} {130503} (\bibinfo {year} {2012})}\BibitemShut {NoStop}%
\bibitem [{\citenamefont {Lucamarini}\ \emph {et~al.}(2018)\citenamefont {Lucamarini}, \citenamefont {Yuan}, \citenamefont {Dynes},\ and\ \citenamefont {Shields}}]{lucamarini2018overcoming}%
  \BibitemOpen
  \bibfield  {author} {\bibinfo {author} {\bibfnamefont {M.}~\bibnamefont {Lucamarini}}, \bibinfo {author} {\bibfnamefont {Z.~L.}\ \bibnamefont {Yuan}}, \bibinfo {author} {\bibfnamefont {J.~F.}\ \bibnamefont {Dynes}},\ and\ \bibinfo {author} {\bibfnamefont {A.~J.}\ \bibnamefont {Shields}},\ }\bibfield  {title} {\bibinfo {title} {Overcoming the rate--distance limit of quantum key distribution without quantum repeaters},\ }\href@noop {} {\bibfield  {journal} {\bibinfo  {journal} {Nature}\ }\textbf {\bibinfo {volume} {557}},\ \bibinfo {pages} {400} (\bibinfo {year} {2018})}\BibitemShut {NoStop}%
\bibitem [{\citenamefont {Xie}\ \emph {et~al.}(2022)\citenamefont {Xie}, \citenamefont {Lu}, \citenamefont {Weng}, \citenamefont {Cao}, \citenamefont {Jia}, \citenamefont {Bao}, \citenamefont {Wang}, \citenamefont {Fu}, \citenamefont {Yin},\ and\ \citenamefont {Chen}}]{xie2022breaking}%
  \BibitemOpen
  \bibfield  {author} {\bibinfo {author} {\bibfnamefont {Y.-M.}\ \bibnamefont {Xie}}, \bibinfo {author} {\bibfnamefont {Y.-S.}\ \bibnamefont {Lu}}, \bibinfo {author} {\bibfnamefont {C.-X.}\ \bibnamefont {Weng}}, \bibinfo {author} {\bibfnamefont {X.-Y.}\ \bibnamefont {Cao}}, \bibinfo {author} {\bibfnamefont {Z.-Y.}\ \bibnamefont {Jia}}, \bibinfo {author} {\bibfnamefont {Y.}~\bibnamefont {Bao}}, \bibinfo {author} {\bibfnamefont {Y.}~\bibnamefont {Wang}}, \bibinfo {author} {\bibfnamefont {Y.}~\bibnamefont {Fu}}, \bibinfo {author} {\bibfnamefont {H.-L.}\ \bibnamefont {Yin}},\ and\ \bibinfo {author} {\bibfnamefont {Z.-B.}\ \bibnamefont {Chen}},\ }\bibfield  {title} {\bibinfo {title} {Breaking the rate-loss bound of quantum key distribution with asynchronous two-photon interference},\ }\href@noop {} {\bibfield  {journal} {\bibinfo  {journal} {PRX Quantum}\ }\textbf {\bibinfo {volume} {3}},\ \bibinfo {pages} {020315} (\bibinfo {year} {2022})}\BibitemShut {NoStop}%
\bibitem [{\citenamefont {Fu}\ \emph {et~al.}(2015)\citenamefont {Fu}, \citenamefont {Yin}, \citenamefont {Chen},\ and\ \citenamefont {Chen}}]{fu2015long}%
  \BibitemOpen
  \bibfield  {author} {\bibinfo {author} {\bibfnamefont {Y.}~\bibnamefont {Fu}}, \bibinfo {author} {\bibfnamefont {H.-L.}\ \bibnamefont {Yin}}, \bibinfo {author} {\bibfnamefont {T.-Y.}\ \bibnamefont {Chen}},\ and\ \bibinfo {author} {\bibfnamefont {Z.-B.}\ \bibnamefont {Chen}},\ }\bibfield  {title} {\bibinfo {title} {Long-distance measurement-device-independent multiparty quantum communication},\ }\href@noop {} {\bibfield  {journal} {\bibinfo  {journal} {Phys. Rev. Lett.}\ }\textbf {\bibinfo {volume} {114}},\ \bibinfo {pages} {090501} (\bibinfo {year} {2015})}\BibitemShut {NoStop}%
\bibitem [{\citenamefont {Cao}\ \emph {et~al.}(2023)\citenamefont {Cao}, \citenamefont {Lu}, \citenamefont {Chai}, \citenamefont {Yu}, \citenamefont {Liang},\ and\ \citenamefont {Wang}}]{cao2023realization}%
  \BibitemOpen
  \bibfield  {author} {\bibinfo {author} {\bibfnamefont {Z.}~\bibnamefont {Cao}}, \bibinfo {author} {\bibfnamefont {Y.}~\bibnamefont {Lu}}, \bibinfo {author} {\bibfnamefont {G.}~\bibnamefont {Chai}}, \bibinfo {author} {\bibfnamefont {H.}~\bibnamefont {Yu}}, \bibinfo {author} {\bibfnamefont {K.}~\bibnamefont {Liang}},\ and\ \bibinfo {author} {\bibfnamefont {L.}~\bibnamefont {Wang}},\ }\bibfield  {title} {\bibinfo {title} {Realization of quantum secure direct communication with continuous variable},\ }\href@noop {} {\bibfield  {journal} {\bibinfo  {journal} {Research}\ }\textbf {\bibinfo {volume} {6}},\ \bibinfo {pages} {0193} (\bibinfo {year} {2023})}\BibitemShut {NoStop}%
\bibitem [{\citenamefont {Shen}\ \emph {et~al.}(2023)\citenamefont {Shen}, \citenamefont {Cao}, \citenamefont {Wang}, \citenamefont {Fu}, \citenamefont {Gu}, \citenamefont {Liu}, \citenamefont {Weng}, \citenamefont {Yin},\ and\ \citenamefont {Chen}}]{shen2023experimental}%
  \BibitemOpen
  \bibfield  {author} {\bibinfo {author} {\bibfnamefont {A.}~\bibnamefont {Shen}}, \bibinfo {author} {\bibfnamefont {X.-Y.}\ \bibnamefont {Cao}}, \bibinfo {author} {\bibfnamefont {Y.}~\bibnamefont {Wang}}, \bibinfo {author} {\bibfnamefont {Y.}~\bibnamefont {Fu}}, \bibinfo {author} {\bibfnamefont {J.}~\bibnamefont {Gu}}, \bibinfo {author} {\bibfnamefont {W.-B.}\ \bibnamefont {Liu}}, \bibinfo {author} {\bibfnamefont {C.-X.}\ \bibnamefont {Weng}}, \bibinfo {author} {\bibfnamefont {H.-L.}\ \bibnamefont {Yin}},\ and\ \bibinfo {author} {\bibfnamefont {Z.-B.}\ \bibnamefont {Chen}},\ }\bibfield  {title} {\bibinfo {title} {Experimental quantum secret sharing based on phase encoding of coherent states},\ }\href@noop {} {\bibfield  {journal} {\bibinfo  {journal} {Science China Physics, Mechanics \& Astronomy}\ }\textbf {\bibinfo {volume} {66}},\ \bibinfo {pages} {260311} (\bibinfo {year} {2023})}\BibitemShut {NoStop}%
\bibitem [{\citenamefont {Amiri}\ \emph {et~al.}(2018)\citenamefont {Amiri}, \citenamefont {Abidin}, \citenamefont {Wallden},\ and\ \citenamefont {Andersson}}]{amiri2018efficient}%
  \BibitemOpen
  \bibfield  {author} {\bibinfo {author} {\bibfnamefont {R.}~\bibnamefont {Amiri}}, \bibinfo {author} {\bibfnamefont {A.}~\bibnamefont {Abidin}}, \bibinfo {author} {\bibfnamefont {P.}~\bibnamefont {Wallden}},\ and\ \bibinfo {author} {\bibfnamefont {E.}~\bibnamefont {Andersson}},\ }\bibfield  {title} {\bibinfo {title} {Efficient unconditionally secure signatures using universal hashing},\ }in\ \href@noop {} {\emph {\bibinfo {booktitle} {Applied Cryptography and Network Security}}},\ \bibinfo {editor} {edited by\ \bibinfo {editor} {\bibfnamefont {B.}~\bibnamefont {Preneel}}\ and\ \bibinfo {editor} {\bibfnamefont {F.}~\bibnamefont {Vercauteren}}}\ (\bibinfo  {publisher} {Springer International Publishing},\ \bibinfo {year} {2018})\ pp.\ \bibinfo {pages} {143--162}\BibitemShut {NoStop}%
\bibitem [{\citenamefont {Wallden}\ \emph {et~al.}(2015)\citenamefont {Wallden}, \citenamefont {Dunjko}, \citenamefont {Kent},\ and\ \citenamefont {Andersson}}]{wallden2015quantum}%
  \BibitemOpen
  \bibfield  {author} {\bibinfo {author} {\bibfnamefont {P.}~\bibnamefont {Wallden}}, \bibinfo {author} {\bibfnamefont {V.}~\bibnamefont {Dunjko}}, \bibinfo {author} {\bibfnamefont {A.}~\bibnamefont {Kent}},\ and\ \bibinfo {author} {\bibfnamefont {E.}~\bibnamefont {Andersson}},\ }\bibfield  {title} {\bibinfo {title} {Quantum digital signatures with quantum-key-distribution components},\ }\href@noop {} {\bibfield  {journal} {\bibinfo  {journal} {Phys. Rev. A}\ }\textbf {\bibinfo {volume} {91}},\ \bibinfo {pages} {042304} (\bibinfo {year} {2015})}\BibitemShut {NoStop}%
\bibitem [{\citenamefont {Comandar}\ \emph {et~al.}(2016)\citenamefont {Comandar}, \citenamefont {Lucamarini}, \citenamefont {Fr{\"o}hlich}, \citenamefont {Dynes}, \citenamefont {Sharpe}, \citenamefont {Tam}, \citenamefont {Yuan}, \citenamefont {Penty},\ and\ \citenamefont {Shields}}]{comandar2016quantum}%
  \BibitemOpen
  \bibfield  {author} {\bibinfo {author} {\bibfnamefont {L.}~\bibnamefont {Comandar}}, \bibinfo {author} {\bibfnamefont {M.}~\bibnamefont {Lucamarini}}, \bibinfo {author} {\bibfnamefont {B.}~\bibnamefont {Fr{\"o}hlich}}, \bibinfo {author} {\bibfnamefont {J.}~\bibnamefont {Dynes}}, \bibinfo {author} {\bibfnamefont {A.}~\bibnamefont {Sharpe}}, \bibinfo {author} {\bibfnamefont {S.-B.}\ \bibnamefont {Tam}}, \bibinfo {author} {\bibfnamefont {Z.}~\bibnamefont {Yuan}}, \bibinfo {author} {\bibfnamefont {R.}~\bibnamefont {Penty}},\ and\ \bibinfo {author} {\bibfnamefont {A.}~\bibnamefont {Shields}},\ }\bibfield  {title} {\bibinfo {title} {Quantum key distribution without detector vulnerabilities using optically seeded lasers},\ }\href@noop {} {\bibfield  {journal} {\bibinfo  {journal} {Nat. Photonics}\ }\textbf {\bibinfo {volume} {10}},\ \bibinfo {pages} {312} (\bibinfo {year} {2016})}\BibitemShut {NoStop}%
\end{thebibliography}
%

\end{document}